\documentclass[useAMS,usenatbib]{mn2e}
\usepackage{txfonts}
\usepackage{graphicx}
\usepackage{natbib}

\def\del#1{{}}

\newcommand{\dd}{\mathrm{d}}
\newcommand{\bra}{\langle}
\newcommand{\ket}{\rangle}
\newcommand{\ltsima}{$\; \buildrel < \over \sim \;$}
\newcommand{\lsim}{\lower.5ex\hbox{\ltsima}}
\newcommand{\gtsima}{$\; \buildrel > \over \sim \;$}
\newcommand{\gsim}{\lower.5ex\hbox{\gtsima}}

\newcommand{\e}{{\rm e}}
\newcommand{\p}{{\rm p}}
\newcommand{\inj}{{\rm inj}}

\newcommand{\W}{{\mathcal W}}

\newcommand{\eps}{\varepsilon}
\newcommand{\vecbf}{\mathbfit}

\newcommand{\ie}{{\textit{i.e.\ }}}

\title[Exploring the magnetized cosmic web]{Exploring the magnetized cosmic web
  through low frequency radio emission}

\author[Battaglia et. al.] 
{N. Battaglia$^{1,2}$\thanks{ e-mail:
    battaglia@astro.utoronto.ca (NB); pfrommer@cita.utoronto.ca
    (CP); sievers@cita.utoronto.ca (JLS); bond@cita.utoronto.ca (JRB); ensslin@mpa-garching.mpg.de (TAE)},
  C. Pfrommer$^2$, J. L. Sievers$^2$, J. R. Bond$^2$,
  T. A. En{\ss}lin$^3$  
  \\$^1$Department of
  Astronomy and Astrophysics, University of Toronto, 50 St. George st.,
  Toronto ON, M5S 3H4, \\$^2$Canadian Institute for Theoretical Astrophysics,
  60 St. George st., Toronto ON, M5S 3H8\\$^3$Max-Planck-Institut f\"ur Astrophysik,
  Karl-Schwarzschild-Stra{\ss}e 1, Postfach 1317, 85741 Garching,
  Germany}

\begin{document}
\pagerange{\pageref{firstpage}--\pageref{lastpage}} \pubyear{2008}
\maketitle
\label{firstpage}

\begin{abstract}
  Recent improvements in the capabilities of low frequency radio
  telescopes provide a unique opportunity to study thermal and
  non-thermal properties of the cosmic web. We argue that the diffuse,
  polarized emission from giant radio relics traces structure
  formation shock waves and illuminates the large-scale magnetic
  field. To show this, we model the population of shock-accelerated
  relativistic electrons in high-resolution cosmological simulations
  of galaxy clusters and calculate the resulting radio synchrotron
  emission. We find that individual shock waves correspond to
  localized peaks in the radio surface brightness map which enables us
  to measure Mach numbers for these shocks. We show that the
  luminosities and number counts of the relics strongly depend on the
  magnetic field properties, the cluster mass and dynamical state. By
  suitably combining different cluster data, including Faraday
  rotation measures, we are able to constrain some macroscopic
  parameters of the plasma at the structure formation shocks, such as
  models of turbulence.  We also predict upper limits for the properties
  of the warm-hot intergalactic medium, such as its temperature and
  density. We predict that the current generation of radio telescopes
  (LOFAR, GMRT, MWA, LWA) have the potential to discover a
  substantially larger sample of radio relics, with multiple relics
  expected for each violently merging cluster.  Future experiments
  (SKA) should enable us to further probe the macroscopic parameters
  of plasma physics in clusters.
\end{abstract}


\section{Introduction and key questions}

The plasma within and between galaxies is magnetized.  Despite many
observational efforts to measure galactic and intergalactic magnetic
fields, their properties and origins are not currently well
understood. The magnetic fields influence the physics of the plasma in
several important ways.  They couple the collisionless charged
particles to a single but complex fluid through the Lorentz force, and
trace dynamical processes in the Universe. Magnetic pressure and
tension mediate forces and provide the plasma with additional
macroscopic degrees of freedom in terms of Alfv{\'e}nic and
magnetosonic waves. They cause the turbulent cascade to become
anisotropic towards smaller scales and suppress transport processes
such as heat conduction and cosmic ray diffusion across the mean
magnetic field. They are essential for accelerating cosmic rays by
providing macroscopic scattering agents which enables diffusive shock
acceleration (first order Fermi process) and through
magneto-hydrodynamic turbulent interactions with cosmic rays leading
to second order Fermi acceleration. They illuminate distant cosmic ray
electron populations by enabling synchrotron emission and tell us
indirectly about violent high-energy astrophysical processes such as
formation shock waves or $\gamma$-ray bursts. The magnetic fields in
spiral galaxies are highly regular, showing alignment with the spiral
arms. They are believed to arise from weak seed fields amplified by
dynamo processes, driven by differential rotation in galactic disks.
The seed fields could have been produced by many sources, ranging from
stellar winds and jets of active galactic nuclei, to plasma
instabilities and battery effects in shock waves, in ionization
fronts, and in neutral gas-plasma interactions. More hypothetical
ideas for the seed field origins invoke primordial generation in early
Universe processes, such as phase transitions during the epoch of
inflation.  In order to understand more about magneto-genesis, we need
to study the least processed plasma possible that still shows some
primordial memory. This points us to the magnetized plasma in
intergalactic space, in particular to the plasma in galaxy
clusters. There, magnetic fields show a smaller degree of ordering
compared to spiral galaxies.  However, their primordial properties may
be masked in clusters because of processing by turbulent gas flows,
driven by galaxy cluster mergers, and the orbits of the member
galaxies.  For an overview on the present observational and
theoretical knowledge the reader is pointed to the review articles by
\citet{1987QJRAS..28..197R,1993A&ARv...4..449W, 1994RPPh...57..325K,
1996ARA&A..34..155B, 1999ARA&A..37...37K, 2001SSRv...99..243B,
2001PhR...348..163G, 2002ARA&A..40..319C, 2002RvMP...74..775W}. This
work aims at closing a gap between theoretically motivated
phenomenological models of large scale magnetic fields and actual
observational non-thermal phenomena associated with them.

Diffuse radio synchrotron emission has already been observed in more than
50 galaxy clusters \citep{2008SSRv..134...93F}.  The emission is
associated with the entire intra-cluster medium (ICM).  The
synchrotron emission process demonstrates the presence of highly
relativistic electrons (cosmic ray electrons, CRe) with a Lorentz
factor typically up to $\gamma \sim 10^4$ and magnetic fields within
the ICM.  The diffuse radio emission can be classified into two
categories: radio halos and radio relics. Giant radio halos are
centrally located, trace the thermal emission and show no sign of
polarization, while radio relics are located at the periphery of
clusters, are polarized and are elongated in appearance. There exist
a number of classes of radio objects that have been referred to over the
years as ``radio relics'' \citep[][ and references
therein]{2004rcfg.proc..335K}. Two of these are associated with
extinct or dying active galactic nuclei (AGN). These either host a
synchrotron cooling radio plasma from a past AGN outburst that created
the radio lobes or are revived ``radio ghosts'' where an aged radio
relic has been re-energized by a merger or an accretion shock
\citep{2001A&A...366...26E}.  

The focus of this paper is on a third type of radio relic emission,
sometimes referred to as radio ``gischt''\footnote{The name ``gischt''
derives from a German word for the crest on top of waves that are
breaking at the shore thus resembling the radio emission of freshly
injected electrons by formation shocks.} \citep{2004rcfg.proc..335K},
that shows diffuse emission on scales up to 1~Mpc.  Diffusive shock
acceleration at structure formation shocks can energize a primary
population of relativistic electrons that emit synchrotron radiation
\citep{1998A&A...332..395E,2001ApJ...562..233M} in a magnetic field
that can be amplified by the post-shock turbulence.  Prominent
examples for this class of radio relics can be seen in Abell 3667
\citep{1997MNRAS.290..577R}, Abell 2256
\citep{1976A&A....52..107B,1978MNRAS.185..607M, 1979A&A....80..201B,
1994ApJ...436..654R, 2006AN....327..553C}, Abell 3376
\citep{2006Sci...314..791B}, and, more recently, Abell 2255
\citep{2008A&A...481L..91P} and Abell 521 \citep{2008A&A...486..347G}.
All galaxy clusters with observed radio relic emission are merging or
show signs of ongoing dynamical activity, but not all dynamically
active galaxy clusters are observed to have relics. This raises the
question of whether diffuse radio emission is a property of a special
subset of clusters, or a universal property, with many relics too
faint to be seen by current telescopes.

From CMB measurements we know that the Universe is composed of 4.6\%
baryonic matter, {\it e.g.} \citet{2008arXiv0803.0547K}. However, when
observing the local Universe ($z<1$), we can account for fewer than
half of these baryons
\citep{2004IAUS..220..227F,2005ApJ...624..555D}. This is known as the
missing baryon problem. The current cosmological paradigm of large
scale structure formation provides a solution to the missing baryon
problem. As the Universe evolves, large scale structure grows from
small density perturbations imprinted during an earlier epoch. In the
hierarchical scenario of structure formation, structure grows from
small to large scales, with baryons flowing in a filamentary web with
clusters at the interstices. The temperature of baryons deviates from
adiabatic cooling associated with the Hubble expansion by increasing
multiple times in discrete steps -- always corresponding to a passage
through a structure formation shock. Before they are shock-heated to
the virial temperatures $kT \sim 1 - 10$~keV of galaxy groups and
clusters, where they can be observed through their thermal
bremsstrahlung emission, they are predicted to reside in the warm-hot
intergalactic medium (WHIM).  Temperatures in the WHIM are in the
range of $10^5\,\mbox{K}<T<10^7\,\mbox{K}$ \citep{1998ApJ...509...56H,
1999ApJ...514....1C, 2001ApJ...552..473D, 2004ApJ...611..642F,
2005ApJ...620...21K}.  We will investigate whether it is possible for
diffuse radio emission associated with these formation shocks to be
used as a tracer of the WHIM boundaries and so indirectly observe the
WHIM.

\begin{figure*}
\begin{center}
 \begin{minipage}[t]{0.55\textwidth}
   \centering{\it \large Mach numbers}
 \end{minipage}
 \hfill
 \begin{minipage}[t]{0.43\textwidth}
   \centering{\it \large Radio Gischt at shocks}
 \end{minipage}
\resizebox{0.53\hsize}{!}{\includegraphics{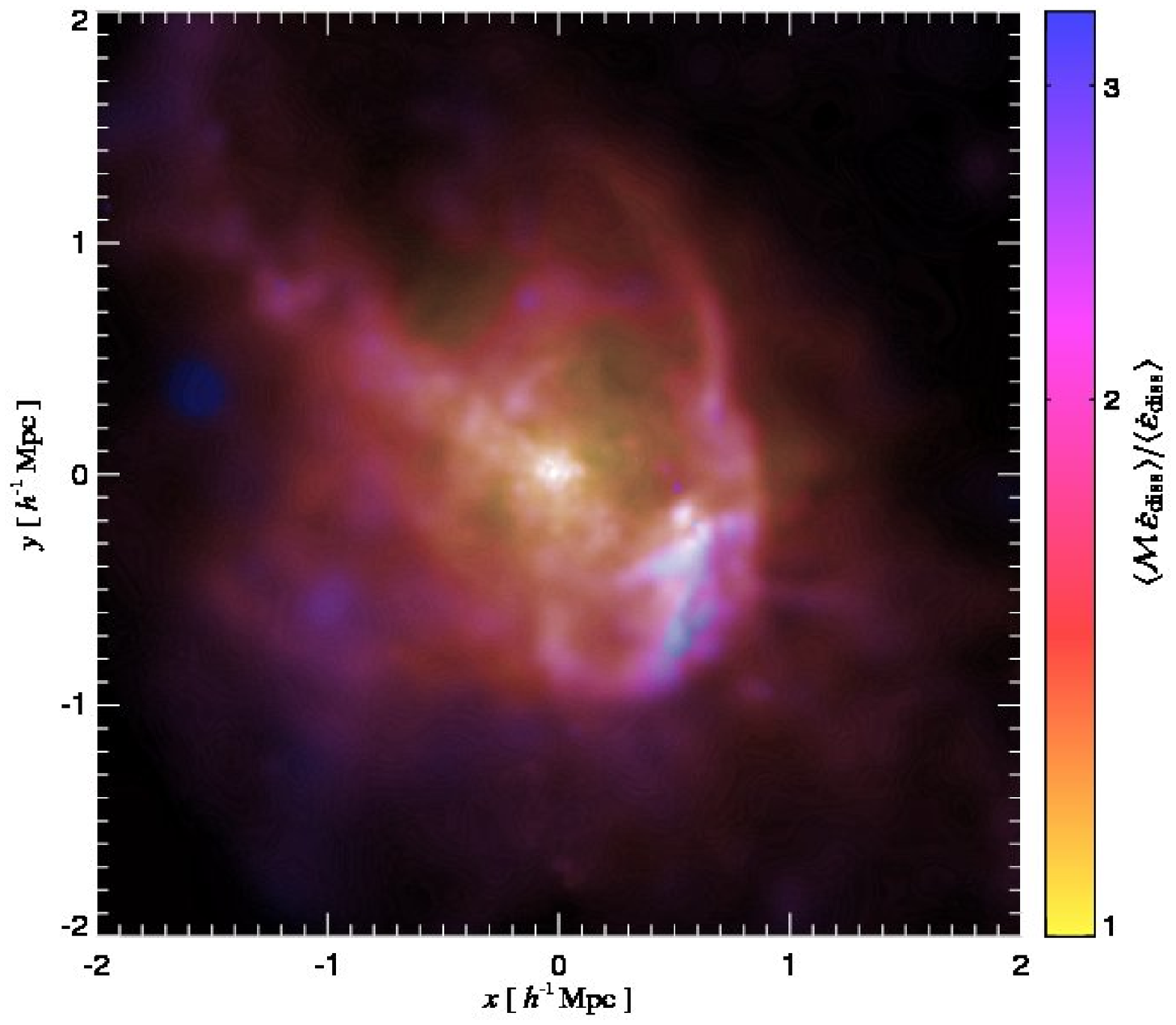}}%
\resizebox{0.46\hsize}{!}{\includegraphics{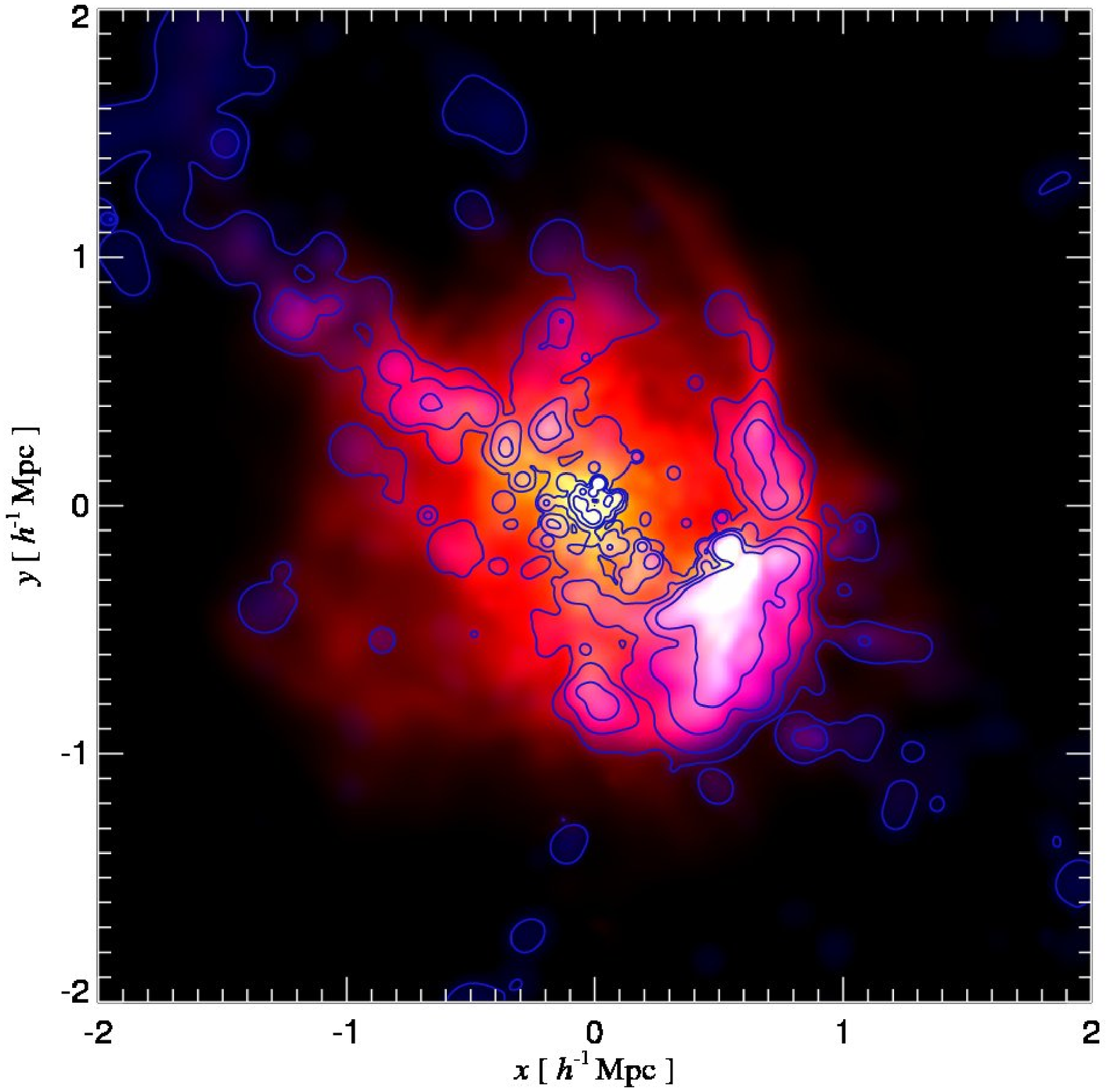}}\\
\end{center}
\caption{Structure formation shocks triggered by a recent merger of a large
  galaxy cluster ($M\simeq 10^{15} h^{-1}\,\rmn{M}_\odot$) dissipate the
  associated gravitational energy. Left: the Mach number of shocks
  weighted by the energy dissipation rate is shown by the colour (while the brightness
  displays the logarithm of the dissipation rate). Right: three-color image 
  of energy dissipation rate at shocks (shown with a color scale ranging from
  black over red to yellow) and radio synchrotron emission at 150 MHz from
  shock-accelerated relativistic electrons (blue and contours with levels
  starting at $7\times 10^{-4}$ mJy arcmin$^{-2}$ and increasing with a factor of
  15, respectively). This ``radio gischt'' emission traces structure formation
  shock waves, highlights the intermittent nature of mass accretion in galaxy
  clusters, and illuminates magnetic fields that are amplified by turbulence
  that can be excited by these shock waves. }
\label{fig:machnumbers}
\end{figure*}

A larger sample size of these diffuse radio sources is required to
delve deep into the details of the non-thermal processes working
within galaxy clusters. However, the combination of low surface
brightness, diffuseness and small dynamical range in the sensitivity
of current radio telescopes makes the detection of this particular
radio emission difficult. With the current capabilities of the
Giant Meter Radio Telescope (GMRT, \citealt{1995JApAS..16..427A}) and
the imminent arrivals of the Low Frequency ARray (LOFAR,
\citealt{2003NewAR..47..405R}), the Murchison Wide-field Array (MWA,
\citealt{2006NewAR..50..173M}) and the Long Wavelength Array (LWA,
\citealt{2005ASPC..345..392K}), and eventually the construction of the
Square Kilometre Array (SKA, \citealt{2004NewAR..48.1119K}), powerful
low frequency radio telescopes are positioned to further increase our
understanding of diffuse radio emission and give us insight into the
following important topics:
\begin{itemize}
\item  the strength and coherence scale of magnetic fields on scales of
  galaxy clusters,
\item the process of diffusive shock acceleration of electrons,
\item the existence and properties of the WHIM,
\item the exploration of observables beyond the thermal cluster
  emission which are sensitive to the dynamical state of the cluster.
\end{itemize}

\indent In the course of this work we will consider how radio relic
emission can shed light on each of these topics.  To do this, we adopt
a simplified model for the shock-accelerated population of electrons.
The key figure illustrating these considerations is shown in
Fig.~\ref{fig:machnumbers}.  In our simulations, we can visualize
properties of structure formation shocks that are triggered by a
recent merger of a large galaxy cluster ($M\simeq 10^{15}
h^{-1}\,\rmn{M}_\odot$) and dissipate the associated gravitational
energy. In the left panel, the shock Mach numbers, weighted by the
energy dissipation rate, are encoded by the colour, while the
brightness displays the logarithm of the dissipation rate. This shows
that most of the energy is dissipated in weak flow shocks internal to
the cluster, while the shock waves become strongest and steepen as
they break at the shallower peripheral potentials of the clusters and
within filaments. The right panel shows a three-color image of of
energy dissipation rate at shocks (shown in red and yellow) and radio
synchrotron emission at 150 MHz from shock-accelerated relativistic
electrons (blue and contours, modeled according to
\citealt{2008MNRAS.tmp..277P}).  This ``radio gischt'' emission traces
structure formation shock waves and highlights the intermittent nature
of mass accretion, in particular along the filament extending from the
cluster center to the upper left of the image and for the giant radio
relic to the lower-right of the cluster.  This radio emission
illuminates magnetic fields that are amplified by magneto-hydrodynamic
instabilities that are associated with these shock waves.  The paper
has been structured accordingly: in Sect.~\ref{sec:methodology} we
outline our methodology; we describe our results and discuss them in
Sects.~\ref{sec:results} and \ref{sec:discussion}; and present our conclusions in
Sect.~\ref{sec:conclusions}.

\section{Methodology}
\label{sec:methodology}

We briefly summarize our procedure. We model the synchrotron emission
by calculating the primary shock-accelerated electron population using
a scheme that is based on the thermal leakage model -- a model that
has been developed in the context of diffusive shock acceleration at
supernova remnants \citep{1981JGZG...50..110E}.  We use a simple
parametrization for the magnetic field.  This lets us quickly scan the
observationally allowed parameter space associated with the
mostly-unknown spatial distribution of shocks on cluster scales and
beyond.  In the post-processing, we search for spatially correlated
synchrotron emission from formation shocks, which represent our
simulated radio relics and study the properties of these relics in the
clusters in our sample. Our aim is to understand how radio observables
can be used to reconstruct the physical properties of radio relics,
which trace structure formation and large scale magnetic fields.  In
the subsequent sections there will be a detailed description of our
simulations and modelling.

\begin{figure*}
\begin{center}
 \begin{minipage}[t]{0.33\textwidth}
   \centering{\it \large Observable parameters}
 \end{minipage}
 \hfill
 \begin{minipage}[t]{0.33\textwidth}
   \centering{\it \large Theoretical parameters}
 \end{minipage}
 \hfill
 \begin{minipage}[t]{0.33\textwidth}
   \centering{\it \large Total emission}
 \end{minipage}
\resizebox{0.33\hsize}{!}{\includegraphics{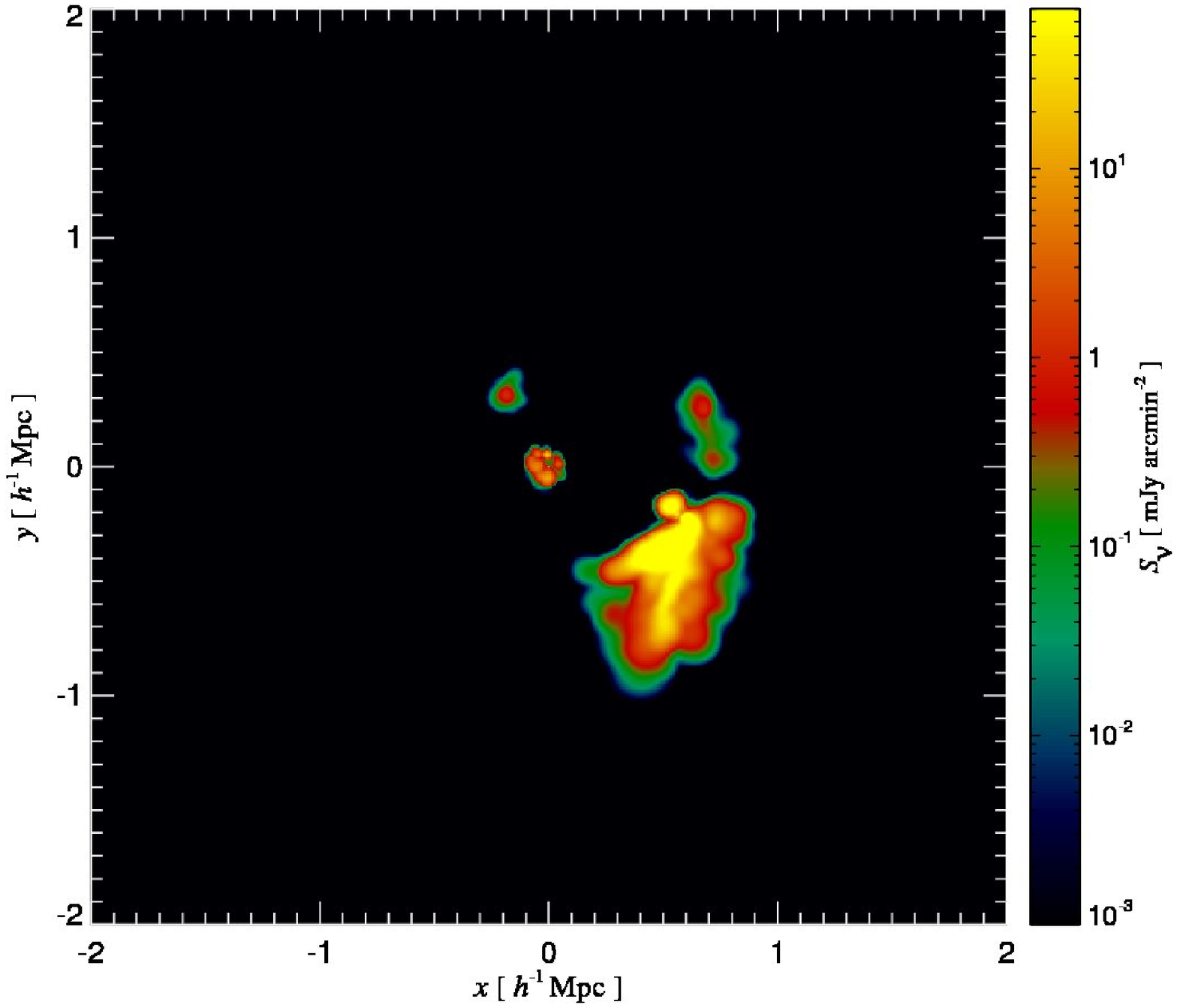}}%
\resizebox{0.33\hsize}{!}{\includegraphics{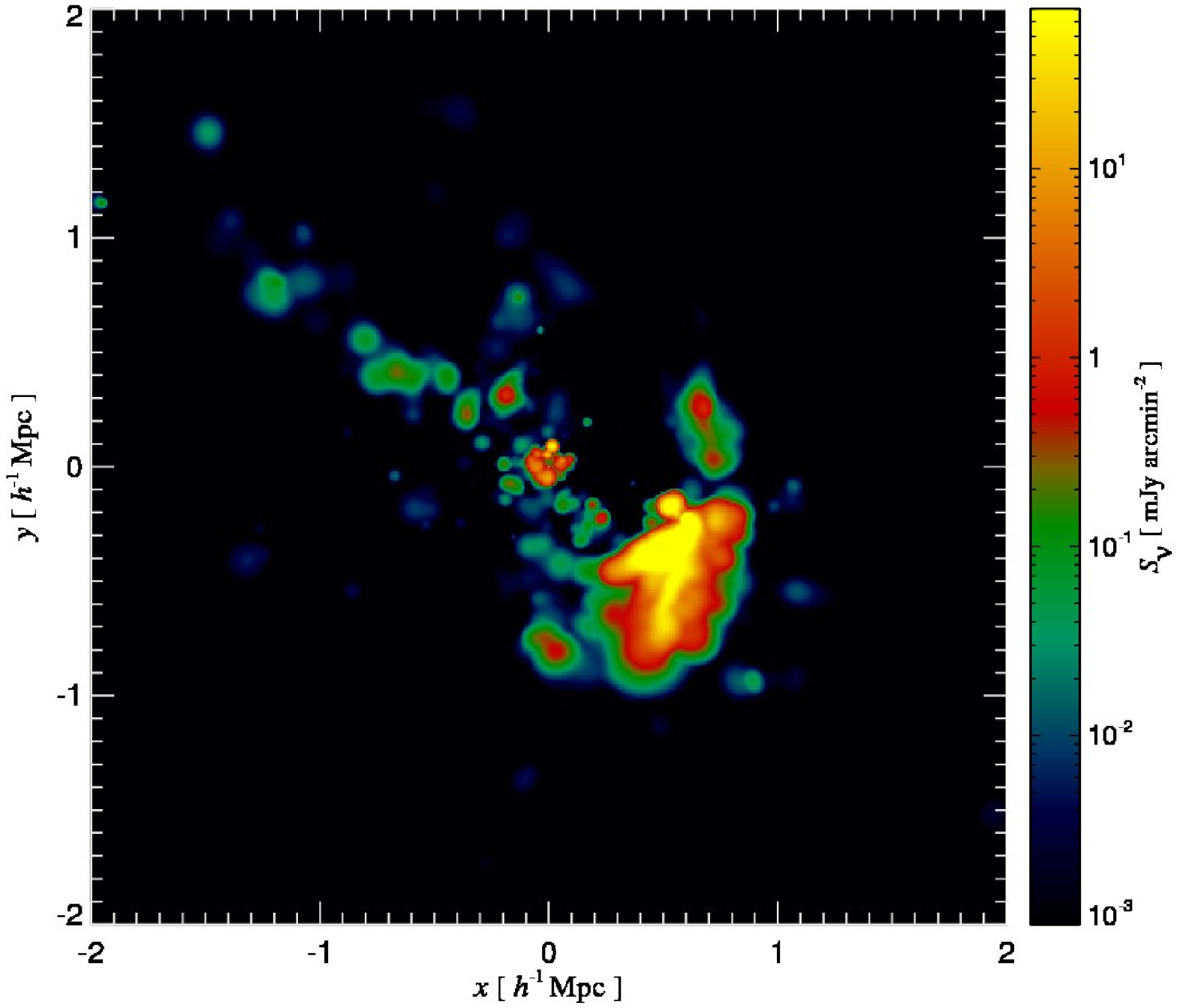}}%
\resizebox{0.33\hsize}{!}{\includegraphics{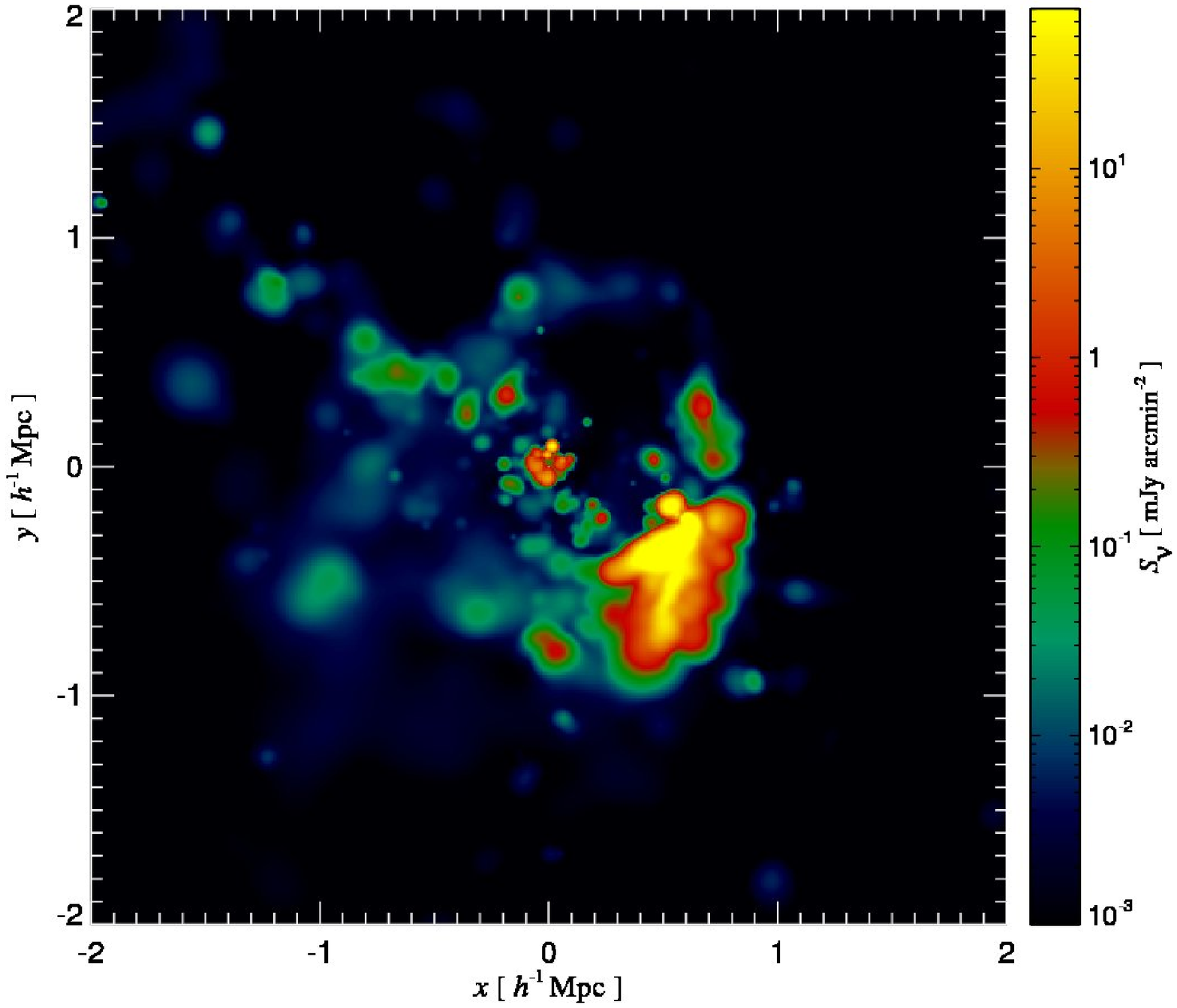}}\\
\end{center}
\caption{Surface brightness emission map for radio relics in the
  simulated cluster g72a   Left to right:  the emission from two sets of relic finder parameters, and the total primary
  radio emission at 150~MHz. Our relic finder groups SPH particles
  using a friends-of-friends algorithm; we additionally require these
  particles to exceed a emissivity threshold (ET). The differences in the images
  illustrate the dependence on ET (left panel: {\em observable parameters} with
  ET = $10^{-43}$ $h^3$ erg s$^{-1}$ Hz$^{-1}$ ster$^{-1}$ cm$^{-3}$, central panel: {\em theoretical
    parameters} ET = $10^{-55}$ $h^3$ erg s$^{-1}$ Hz$^{-1}$ ster$^{-1}$ cm$^{-3}$). The central map only
  lacks surface brightness at the level of $10^{-3}$ compared to the total
  primary emission. Both relic emission maps to the left contain more than 99\%
  of the total flux from the total primary emission map.}
\label{fig:Relicfinder}
\end{figure*}

\subsection{Adopted cosmology and simulated cluster sample} 

Our work is based on high resolution smoothed particle hydrodynamics
(SPH) simulations of galaxy clusters \citep[minimum gas mass
resolution $\sim 8\times 10^9 h^{-1} \mathrm{M}_{\sun}$ for more
details, cf.][]{2007MNRAS.378..385P, 2008MNRAS.tmp..277P} using the
`zoomed initial conditions' technique \citep{1993ApJ...412..455K}
that were selected from a low resolution dark matter only
simulation \citep{2001MNRAS.328..669Y} with a box size of 479 $h^{-1}$
Mpc.  They were carried out using a modified version of the massively
parallel tree SPH code GADGET-2 \citep{2005MNRAS.364.1105S}. The
simulations of the galaxy clusters were performed in a ``Concordance''
cosmology model, $\Lambda$CDM with cosmological parameters of:
$\Omega_{\mathrm{m}}$= $\Omega_{\mathrm{DM}}$ + $\Omega_{\mathrm{b}}$
= 0.3, $\Omega_{\mathrm{b}}$ = 0.039, $\Omega_{\Lambda}$ = 0.7, $h$ =
0.7, $n_s$ = 1 and $\sigma_8$ = 0.9. Here, $\Omega_\rmn{m}$ refers to
the total matter density in units of the critical density today,
$\rho_\rmn{crit} = 3 H_0^2 / (8 \upi G)$. $\Omega_\rmn{b}$ and
$\Omega_\Lambda$ denote the densities of baryons and the cosmological
constant at the present day. The Hubble constant at the present day is
parametrized as $H_0 = 100\,h \mbox{ km s}^{-1} \mbox{Mpc}^{-1}$,
while $n_s$ denotes the spectral index of the primordial
power-spectrum, and $\sigma_8$ is the {\em rms} linear mass
fluctuation within a sphere of radius $8\,h^{-1}$Mpc extrapolated to
$z=0$.

The simulations include a prescription for radiative cooling, star formation,
supernova feedback and a formalism for detecting structure formation shocks and
measuring the associated shock strengths, i.e. the Mach numbers
\citep{2006MNRAS.367..113P}.  Radiative cooling was computed assuming an
optically thin gas of primordial composition (mass-fraction of $X_\rmn{H} =
0.76$ for hydrogen and $1-X_\rmn{H} = 0.24$ for helium) in collisional
ionisation equilibrium, following \citet{1996ApJS..105...19K}. We also
included heating by a photo-ionising, time-dependent, uniform ultraviolet (UV)
background expected from a population of quasars \citep{1996ApJ...461...20H},
which reionises the Universe at $z \simeq 6$. Star formation is treated using
the hybrid multiphase model for the interstellar medium introduced by
\citet{2003MNRAS.339..289S}. In short, the ISM is pictured as a two-phase fluid
consisting of cold clouds that are embedded at pressure equilibrium in an
ambient hot medium.

The cluster sample is displayed in Table~\ref{tab:cluster}. From this sample the
cluster g72a was chosen for detailed analysis of the properties of radio relics
since it is a relatively large (with a mass $M\simeq 10^{15} \rmn{M}_\odot$)
post-merging cluster, similar to the Coma cluster. Additionally, it hosts the
brightest radio relic in the entire sample. This relic resembles already
observed ones.

\begin{table}
\begin{center}
\caption{Cluster sample considered in this paper.}
\begin{tabular}{l|l|c|c|c}
  \hline
  cluster &  dynamical & $M_{200}^b$ & $R_{200}^b$ & $kT_{200}^c$ \\
  name & state$^a$& [$h^{-1}\mathrm{M}_{\sun} $]& [$h^{-1}$ Mpc] & [keV]\\
  \hline
g8a  & CC    & $1.8\times 10^{15}$ & 2.0~  & 13.1 \\
g1a  & CC    & $1.3\times 10^{15}$ & 1.8~  & 10.6 \\
g72a & PostM & $1.1\times 10^{15}$ & 1.7~  & 9.4  \\
g51  & CC    & $1.1\times 10^{15}$ & 1.7~  & 9.4  \\
                                               
g1b  & M     & $3.7\times 10^{14}$ & 1.2~  & 4.7  \\
g72b & M     & $1.5\times 10^{14}$ & 0.87 & 2.4  \\
g1c  & M     & $1.4\times 10^{14}$ & 0.84 & 2.3  \\
g1d  & M     & $9.2\times 10^{13}$ & 0.73 & 1.7  \\
                                               
g676 & CC    & $8.8\times 10^{13}$ & 0.72 & 1.7  \\
g914 & CC    & $8.5\times 10^{13}$ & 0.71 & 1.6  \\
  \hline
\label{tab:cluster}
\end{tabular}
\end{center}
  $^a$ The dynamical state has been classified through a combined criterion
  invoking a merger tree study and the visual inspection of the X-ray
  brightness maps. The labels for the clusters are M--merger, PostM--post
  merger (slightly elongated X-ray contours, weak cool core region
  developing), CC--cool core cluster with extended cooling region (smooth X-ray
  profile). $^b$ The virial mass and radius are related by $M_\Delta(z) = \frac{4}{3}
  \pi\, \Delta\, \rho_\rmn{crit}(z) R_\Delta^3 $, where $\Delta=200$ denotes a
  multiple of the critical overdensity $\rho_\rmn{crit}(z) = 3 H (z)^2/ (8\pi
  G)$. $^c$ The virial temperature is defined by $kT_\Delta = G M_\Delta \, \mu\,
  m_\p / (2 R_\Delta)$, where $\mu$ denotes the mean molecular weight.
\end{table}

\subsection{Realization of magnetic fields}
\label{sec:mag-model}

Current SPH implementations that are capable of following the
magneto-hydrodynamics (MHD) of the gas are presently still fraught with
numerical and physical difficulties, in particular when following dissipative
gas physics \citep{1999A&A...348..351D, 2005JCAP...01..009D,
  2004MNRAS.348..139P, 2005MNRAS.364..384P}. Hence we apply a parametrization in
the post-processing of our completed simulations in order to determine the
strength and morphology of the magnetic field \citep{2008MNRAS.tmp..278P}.
Secondly, the parametrization approach provides us with the advantage of
exploring the parameter space of our magnetic field description more
efficiently, since we are not required to re-simulate when we alter the
ab-initio unknown magnetic field parameters.  We have chosen a simple scaling
model for the magnetic field of
\begin{equation}
  \varepsilon_B = \varepsilon_{B_0} 
  \left(\frac{\varepsilon_{\mathrm{th}}}{\varepsilon_{\mathrm{th}_0}} 
  \right)^{2\alpha_{B}}.
\label{eq:Bscale}
\end{equation}
Our independent model parameters are the magnetic decline $\alpha_{B}$,
and the magnetic core energy density $\varepsilon_{B_0}$. The
thermal energy density $\eps_\rmn{th}$ is measured in units of its central
energy density $\eps_\rmn{th_0} = 3 P_{\mathrm{th}_0}/2$, which we
calculate by
fitting a modified $\beta$-model (Eqn. \ref{eq:BetaPro}) to the radial pressure
profiles of our clusters.  We first remove the over-cooled core (see
Sect.~\ref{sec:removal}),
\begin{equation}
  P(r) = P_{\mathrm{th}_0}\left(1+\left(\frac{r}{r_c}\right)\right)^{-3\beta}.
\label{eq:BetaPro}
\end{equation}
We found that our modified $\beta$-model provides a better fit to the pressure
profiles than the usually adopted spherically symmetric King profiles, i.e.~a
$\beta$-model \citep{1978A&A....70..677C}. 


This parametrization (Eqn.~\ref{eq:Bscale}) was motivated by
non-radiative SPH MHD simulations \citep{1999A&A...348..351D} and
radiative adaptive mesh refinement MHD simulations
\citep{2008A&A...482L..13D} of the formation of galaxy clusters in a
cosmological setting.  Rather than applying a scaling with the gas
density as those simulations suggest, we chose the energy density of
the thermal gas.  Current cosmological radiative simulations (that do
not include feedback from AGN) over-cool the centres of clusters,
giving an overproduction of stars, enhanced central gas densities, and
lower central temperatures than are seen in X-ray
observations. \footnote {Recently, \citet{2008MNRAS.tmp..695S} found
that including cosmic rays from AGN in SPH simulations can solve the
over-cooling problem while providing excellent agreement of the gas
fraction and the inner temperature profile.}  In contrast, the thermal
energy density of the gas is well-behaved in simulations.
Observationally, the parametrization (Eqn.~\ref{eq:Bscale}) is
consistent with statistical studies of Faraday rotation measure maps
\citep{2005A&A...434...67V}.  Theoretically, the growth of magnetic
field strength is determined through turbulent dynamo processes that
will saturate at a field strength determined by the strength of the
magnetic back-reaction \citep[{\it e.g.} ][]{2003PhRvL..90x5003S,
2006PhPl...13e6501S} and is typically a fraction of the turbulent
energy density.  The turbulent energy density should be related to the
thermal energy density, thus motivating our model theoretically.  The
parameter $\varepsilon_{B_0}$ is constrained by past measurements of
magnetic fields within clusters and is chosen such that $B_0 =
[\varepsilon_{B_0}8\pi]^{1/2}$ to be on the order of a few $\umu$G
\citep{2006A&A...460..425G, 2007MNRAS.382...67T,
2008A&A...483..699G}. The parameters explored in our model are shown
in Table~\ref{tab:bpar}.

\begin{table}
\begin{center}
\caption{Magnetic field parameters in various combinations. We define
  our standard magnetic field parameters to be $\alpha_{B}$ =
  0.5, $B_0$ = 5$\umu$G and $\nu$ = 150MHz. These parameters are used
  throughout the paper unless otherwise stated.}
\begin{tabular}{|c|c|}
\hline
 $\alpha_{B}$ &  $B_0$ [$\umu$G]\\
magnetic decline & core magnetic field strength \\
\hline
0.3 & 2.5 \\
0.5 &  5.0 \\
0.7 & 10.0 \\
0.9 & - \\
\hline
\label{tab:bpar}
\end{tabular}
\end{center}
\end{table}

To predict the polarization angle and Faraday rotation measure in our
simulations, we need to model the magnetic morphology of the ICM.  We follow
\citet{1991MNRAS.253..147T} in order to create individual components of the
magnetic vector field that obey a given power spectrum. The details of the
magnetic field structure within the ICM are still unknown. There have been
measurements of magnetic correlations from Faraday rotation measure (RM) maps
which are however limited by the finite window size of radio lobes and hence
only constrain the spectrum on smaller scales. These measurements suggest that
the fields are tangled with a Kolmogorov/Oboukhov-type power spectrum for
coherence lengths of approximately 10 kpc scales and smaller
\citep{2003A&A...412..373V,2005A&A...434...67V,2008A&A...483..699G}. It has been
argued that shallower magnetic field power spectra allow for longer coherence
lengths on the order $100$ kpc \citep{2004A&A...424..429M,
  2006A&A...460..425G}. On the other hand, a Fourier analysis of XMM-Newton
X-ray data reveals the presence of a scale-invariant pressure fluctuation
spectrum in the range between 40 and 90 kpc and is found to be well described by
a projected Kolmogorov/Oboukhov-type turbulence spectrum
\citep{2004A&A...426..387S}. Assuming that the growth of the magnetic field
strength is determined through turbulent dynamo processes suggests a similar
spectrum of magnetic and hydrodynamic turbulence
\citep[{\it e.g.} ][]{2003PhRvL..90x5003S}.

We model the components of the magnetic field, $B_i$, as random
Gaussian fields.  We use a Kolmogorov power spectrum on scales smaller than
the coherence length, and a flat (white-noise) power spectrum on
larger scales.  All three
components of the magnetic field are treated independently, which ensures that the
final distribution of $\vecbf{B} (\vecbf{r})$ has random phases.  After mapping
our SPH Lagrangian energy density distribution of the thermal gas onto
a 3D grid (cf. Appendix Eqn.~\ref{eq:3dproj}), these realizations of the
magnetic field are then scaled such that the magnetic energy density obeys our
assumed scaling given by Eqn.~\ref{eq:Bscale}.  To ensure $\nabla \cdot
\vecbf{B} = 0$, we apply a divergence cleaning procedure to our fields in
Fourier space \citep{1998ApJS..116..133B}:
\begin{equation}
\tilde{B}_i(\vecbf{k}) = 
\sum_{j=1}^3 \left( \delta_{i,j} - \frac{k_i k_j}{k^2} \right) 
\tilde{B}_j(\vecbf{k}).
\label{eq:DivB}
\end{equation}
Applying this procedure to our Gaussian random field removes a third of the
magnetic energy. Thus, we re-normalize $\vecbf{B}$ to conserve the magnetic
energy.

\subsection{Cosmic ray electrons and synchrotron emission}
\label{sect:CRsync}

Collisionless cluster shocks are able to accelerate ions and electrons in the
high-energy tail of their Maxwellian distribution functions through diffusive
shock acceleration \citep[for reviews see][]{1983RPPh...46..973D,
  1987PhR...154....1B, 2001RPPh...64..429M}.  Neglecting non-linear shock
acceleration and cosmic ray modified shock structure, the process of diffusive
shock acceleration uniquely determines the spectrum of the freshly injected
relativistic electron population in the post-shock region that cools and finally
diminishes as a result of loss processes. The radio synchrotron emitting
electron population cools on such a short time scale $\tau_\rmn{sync} <
10^8$~yrs (compared to the very long dynamical time scale $\tau_\rmn{dyn} \sim
1$~Gyr) that we can describe this by instantaneous cooling. In this
approximation, there is no steady-state electron population and we would have to
convert the energy from the electrons to inverse Compton (IC) and synchrotron
radiation. Instead, we introduce a virtual electron population that lives in the
SPH-broadened shock volume only; this is defined to be the volume where energy
dissipation takes place. Within this volume, which is co-moving with the shock, we
can use the steady-state solution for the distribution function of
relativistic electrons and we assume no relativistic electrons in the post-shock
volume, where no energy dissipation occurs. Thus, the cooled CR electron
equilibrium spectrum can be derived from balancing the shock injection with the
IC/synchrotron cooling: above a GeV it is given by
\begin{equation}
\label{eq:f_eq}
f_\e(E) = C_\e\, E^{-\alpha_\e}, \quad
C_\e \propto \frac{\rho}
{\eps_B + \eps_\rmn{ph}}
\end{equation}
Here, $\alpha_\e = \alpha_\inj + 1$ is the spectral index of the equilibrium
electron spectrum and $\eps_\rmn{ph}$ denotes the photon energy density,
taken to be that of CMB photons. A more detailed description of our approach can
be found in \citet{2008MNRAS.tmp..278P}. The synchrotron emissivity $j_{\nu}$
for a power-law spectrum of CRes scales as
\begin{equation}
j_{\nu} \propto C_{\e} B^{\alpha_{\nu} + 1} \nu^{-\alpha_{\nu}},
\label{eq:jnu}
\end{equation}
where $\alpha_{\nu} = (\alpha_{\mathrm{e}} - 1)/2$. A line-of-sight
summation of $j_{\nu}$ yields the radio surface brightness,
$S_{\nu}$. The surface brightness are provided in units of $h = 0.7$
to simplify comparison with observations.

\subsection{Finding radio relics}
\label{sec:relicfinder}

\begin{figure*}
\begin{center}
 \begin{minipage}[t]{0.33\textwidth}
   \centering{\it \large Dirty map}
 \end{minipage}
 \hfill
 \begin{minipage}[t]{0.33\textwidth}
   \centering{\it \large Clean map}
 \end{minipage}
 \hfill
 \begin{minipage}[t]{0.33\textwidth}
   \centering{\it \large Observable parameters}
 \end{minipage}
\resizebox{0.33\hsize}{!}{\includegraphics{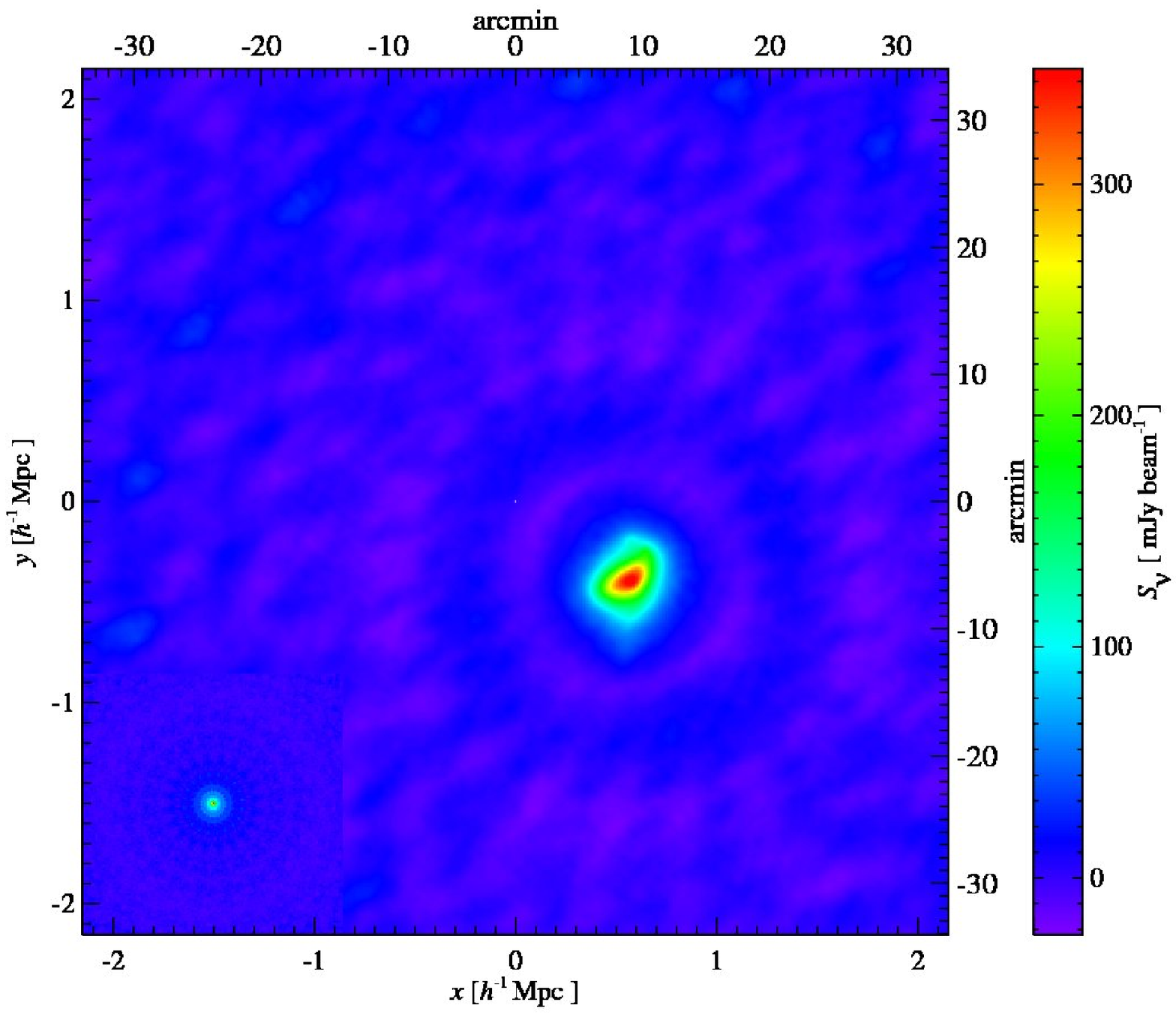}}%
\resizebox{0.33\hsize}{!}{\includegraphics{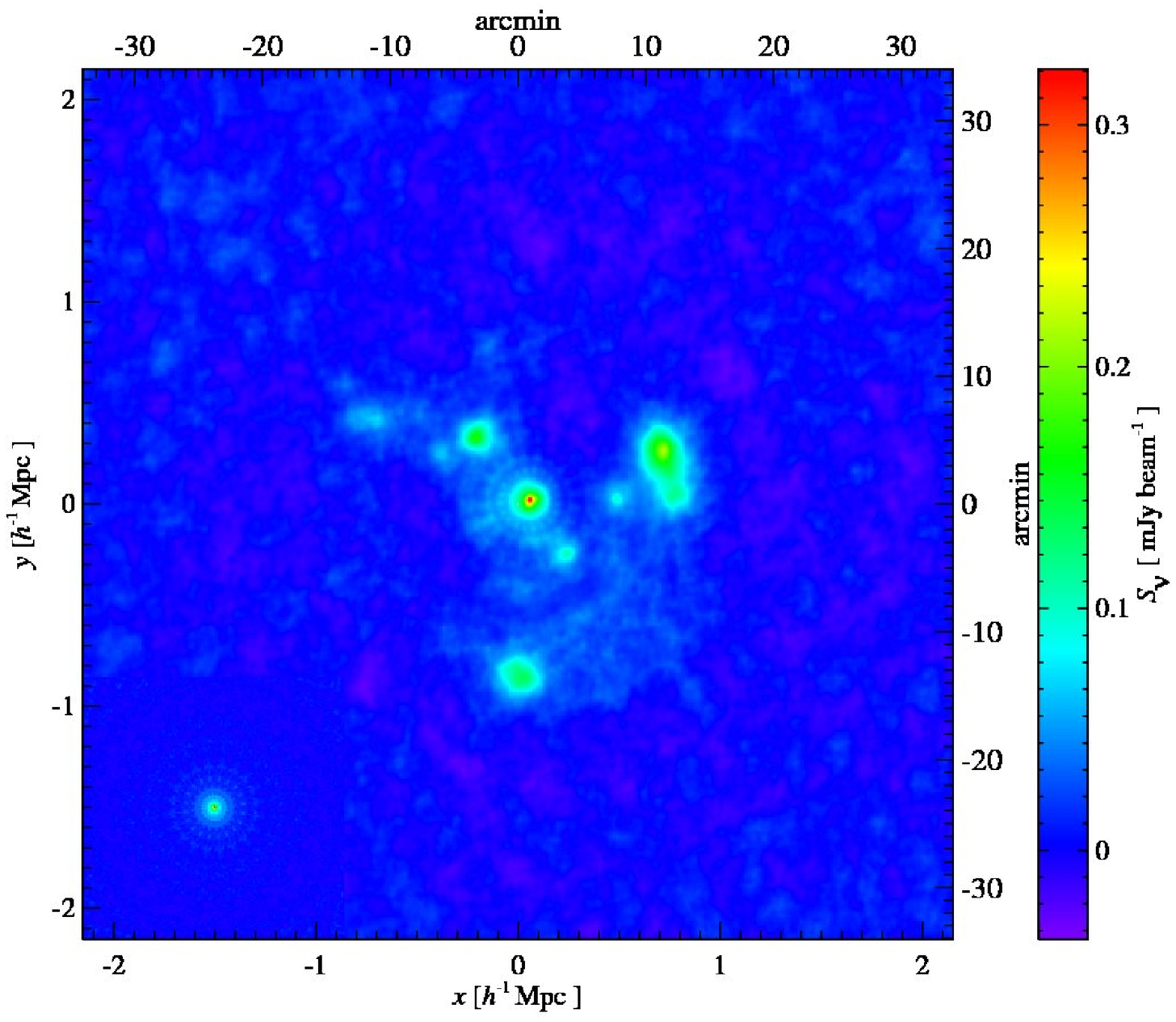}}%
\resizebox{0.33\hsize}{!}{\includegraphics{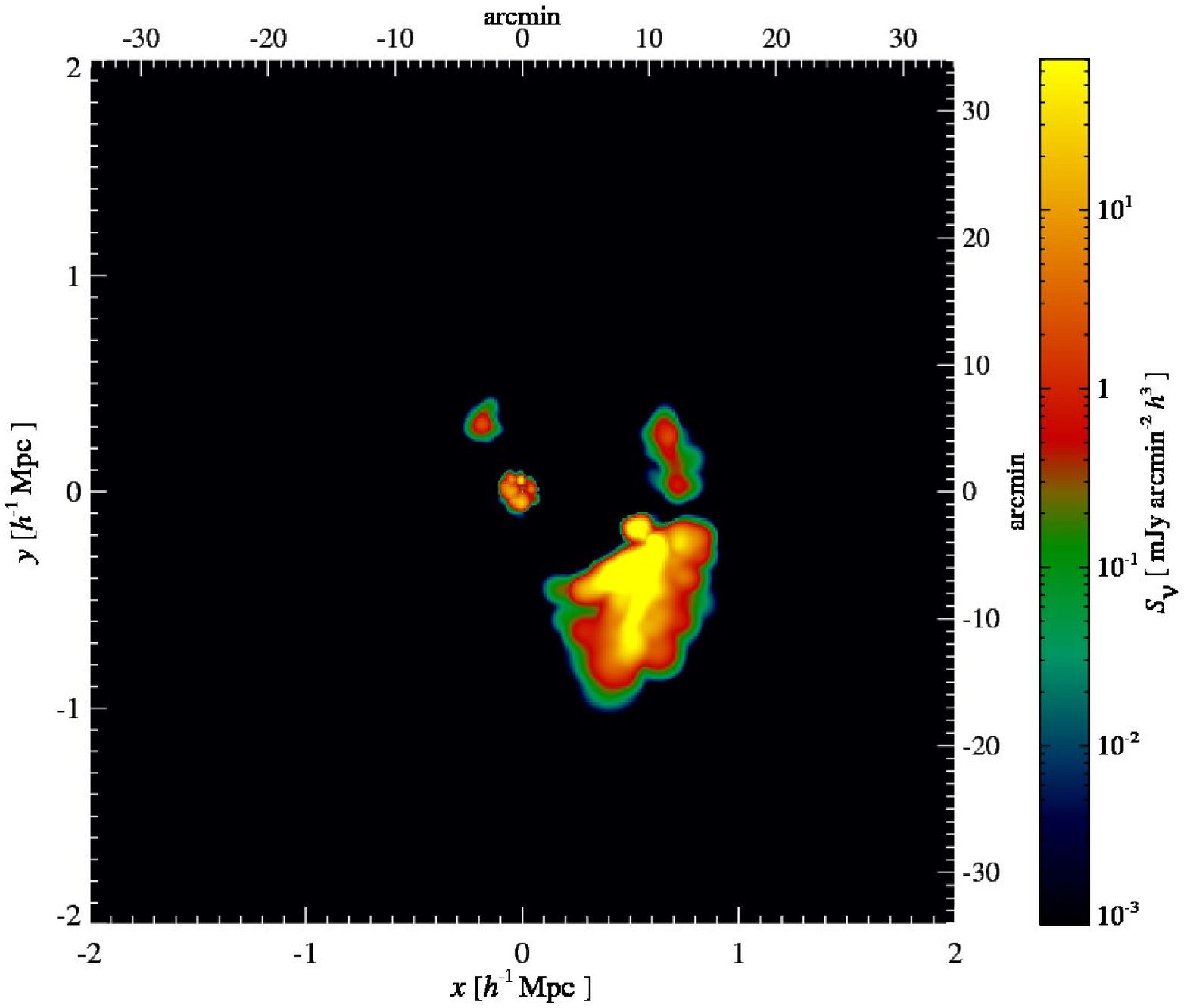}}
\end{center}
\caption{Left two panels: simulated GMRT map of the cluster g72a at z=0.05
  (similar to the Coma cluster), with the reconstructed beam in the
  bottom-left corner.  The left panel is the 'dirty' map and
  the right panel is the 'clean' map, where we removed the brightest relic found
  in the dirty map, mimicking the cleaning procedure of radio maps.  Right
  panel: surface emission map of the relics of g72a using our calibrated set of
  {\em observable parameters} shown with a logarithmic colour scale. Note that
  we reproduce the relic emission that can principally be detected by GMRT after
  applying a cleaning procedure to the compact and diffuse radio emitting
  sources.}
\label{fig:jon}
\end{figure*}

In our search for radio relics in the simulated clusters, we have
modified a friends-of-friends (FOF) \citep{1983ApJS...52...61G}
algorithm so that it groups together
connected radio synchrotron emission in 3D.  This relic
finder works in a manner similar to a FOF finder except we have
introduced the additional criterion of an emission threshold which the
SPH gas particle are required to exceed before being assigned into a
group. Thus, our algorithm depends on three internal parameters which
determine the groups of particles that are designated relics: the
linking length, emissivity threshold, and minimum number of
particles (Fig. \ref{fig:Relicfinder}). The linking length (LL) is the parameter which controls the
maximum distance ($d_{\mathrm{max}}$) between two particles that can
still be considered neighbours,
\begin{equation}
d_{\mathrm{max}} = LL \left[\frac{\bra M_{\rmn{DM}}\ket}
{\Omega_\rmn{DM}\,\rho_{\mathrm{crit}}}\right]^{1/3},
\label{eq:LL}
\end{equation}
where $\rho_\rmn{DM} = \rho_{\mathrm{crit}} \Omega_{\mathrm{DM}}$ is the mass
density of dark matter and $\bra M_{\rmn{DM}}\ket$\footnote{Note that the
  quantity in the brackets is equivalent to the ratio of $\bra M_{\rmn{b}}\ket
  / \rho_\rmn{b}$ except that the baryonic phase consists of gas and stars.} is
the average mass of our dark matter particles.  The linking length and the
emission threshold parameters have degenerate effects on the resulting groups
of particles. Through inspection, we have chosen to fix the linking length
value of 0.2 resulting in $d_{\mathrm{max}}= 50$~kpc and vary the emission
threshold.  A minimum particle value of 32 regulates possible SPH shot noise
and allows for smaller structures to be included in the relic catalogue. The
final parameter in our relic finder, which we have chosen to vary is the
emission threshold. We compute the synchrotron emissivity of all the particles
using Eqn.~\ref{eq:jnu} and compare it to the emission threshold.  The sets of
grouped particles for each of our clusters are our relic catalogues that form
the basis of our study.

\subsubsection{Determination of the emission threshold}
\label{sec:cal}

We tailored the calibration of the emission threshold in the relic
finder to two cases: firstly, so that we would find relics observable
by GMRT and LOFAR; secondly, so that we could study the complete
picture that might be achievable with future radio telescopes such as
SKA, by pushing the emission threshold back to the limit of our
simulations. For this procedure, we left the magnetic field parameters
unchanged. (We used our standard magnetic field parameters
cf. Table~\ref{tab:bpar}.)

We simulate the visibilities and maps from the relics that GMRT would
observe. We use the GMRT primary beam and antenna positions projected
against the zenith (\ie ignoring the z-component of the antenna
positions).  For simplicity, we approximate the continuous UV tracks
for each baseline by circles in the UV plane with measurements
18${\degr}$ apart.  We observe cluster g72a surface emission at a
redshift of 0.05 (cf.{\ }Fig.~\ref{fig:jon}) and make a `dirty' radio
map, the Fourier transform of the visibilities.  We used an
integration time of 2.5 minutes with a sensitivity of 0.2
$\mathrm{mJy}\sqrt{\mathrm{hr}}/\mathrm{beam}$ at the frequency
$\nu=150$~MHz in simulating the visibilities.

To approximate GMRT's dynamical range, we
modelled a simple cleaning procedure by removing the brightest relic
in the 'dirty' map from the total surface emission map and
re-simulated the GMRT detections -- resulting in our `cleaned' radio
map. We compared these maps to surface brightness maps of different
relic catalogues where we varied our emission threshold
(cf. Fig. \ref{fig:jon}). The emission threshold which reproduced the
simulated images most accurately was $10^{-43}$ $h^3$ erg s$^{-1}$ Hz$^{-1}$ ster$^{-1}$ cm$^{-3}$
with the linking length and minimum number of
particles already fixed. These parameters are referred throughout this
paper as {\em observable parameters}.

The choice of a second emission threshold is related to the peak of
the emissivity distribution function which is determined by the mass
resolution of the SPH particles in the simulations. We found the peak
emissivity to be at $j_{\nu}$ = $10^{-55}$ $h^3$ erg s$^{-1}$
Hz$^{-1}$ ster$^{-1}$ cm$^{-3}$. We find that changing $j_{\nu}$ by
six orders of magnitude does not change the number of relics in a
significant way (cf. Appendix \ref{sec:Jontest}), making the
difference between peak emissivity and the {\em observable parameters}
reasonable. Together with the linking length and minimum number of
particle parameters stated above, these parameters are referred to as
{\em theoretical parameters}. Since the emissivity scales with
frequency, the emission threshold must scale with frequency as
well. The emission threshold (ET) scaling is fixed at our reference
frequency $\nu_0 = 150$~MHz,
\begin{equation}
ET = ET_0 \left( \frac{\nu}{\nu_0}\right)^{-1},
\end{equation}
and $ET_0$ adopts the values quoted above for both {\em observational} and {\em
  theoretical parameters}.

In summary, the {\em observable parameters} were chosen to produce relic
catalogues resembling the ones obtainable from current or near future
observations, whereas the {\em theoretical parameters} lead to hypothetical
catalogues only obtainable with a perfect 3-d tomography of the medium which
may find application with the future radio interferometer SKA (cf. Table \ref{tab:emis}).

\begin{table}
\begin{center}
\caption{Parameters chosen for the relic finder.}
\begin{tabular}{|c|c|c|c|}
\hline
 parameter & linking  & minimum number  & emission threshold\\
  name  &   length   & of particles & [erg s$^{-1}$ Hz$^{-1}$ ster$^{-1}$ cm$^{-3}$]\\
\hline
{\em observable} & 0.2 & 32 & 10$^{-55}$ \\
{\em theoretical}  &  0.2 & 32 & 10$^{-43}$ \\
\hline
\label{tab:emis}
\end{tabular}
\end{center}
\end{table}

\begin{figure*}
\begin{center}
 \begin{minipage}[t]{0.495\textwidth}
   \centering{\it \large Observable parameters}
 \end{minipage}
 \hfill
 \begin{minipage}[t]{0.495\textwidth}
   \centering{\it \large Theoretical parameters}
 \end{minipage}
\resizebox{0.5\hsize}{!}{\includegraphics{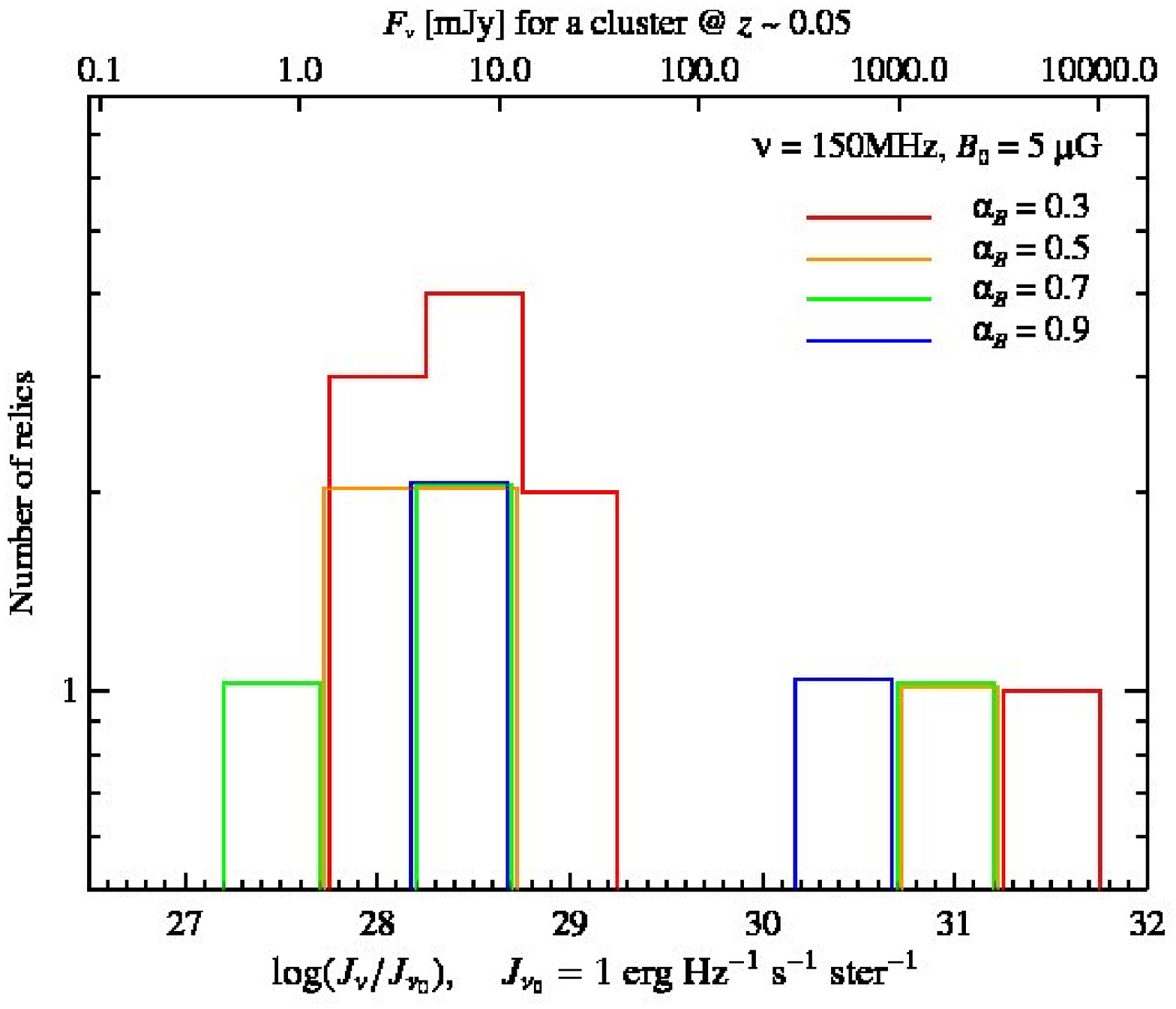}}%
\resizebox{0.5\hsize}{!}{\includegraphics{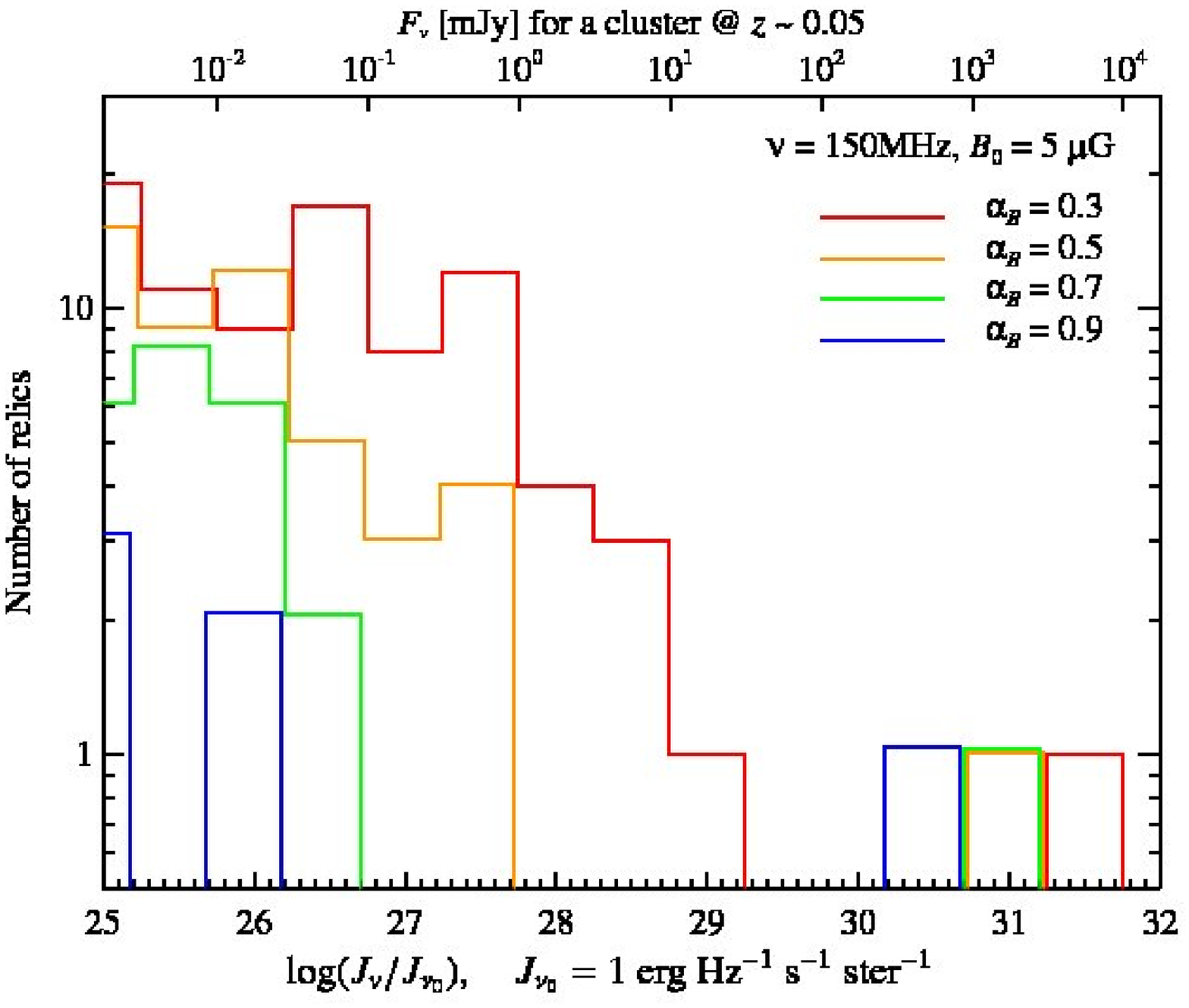}}\\
\resizebox{0.5\hsize}{!}{\includegraphics{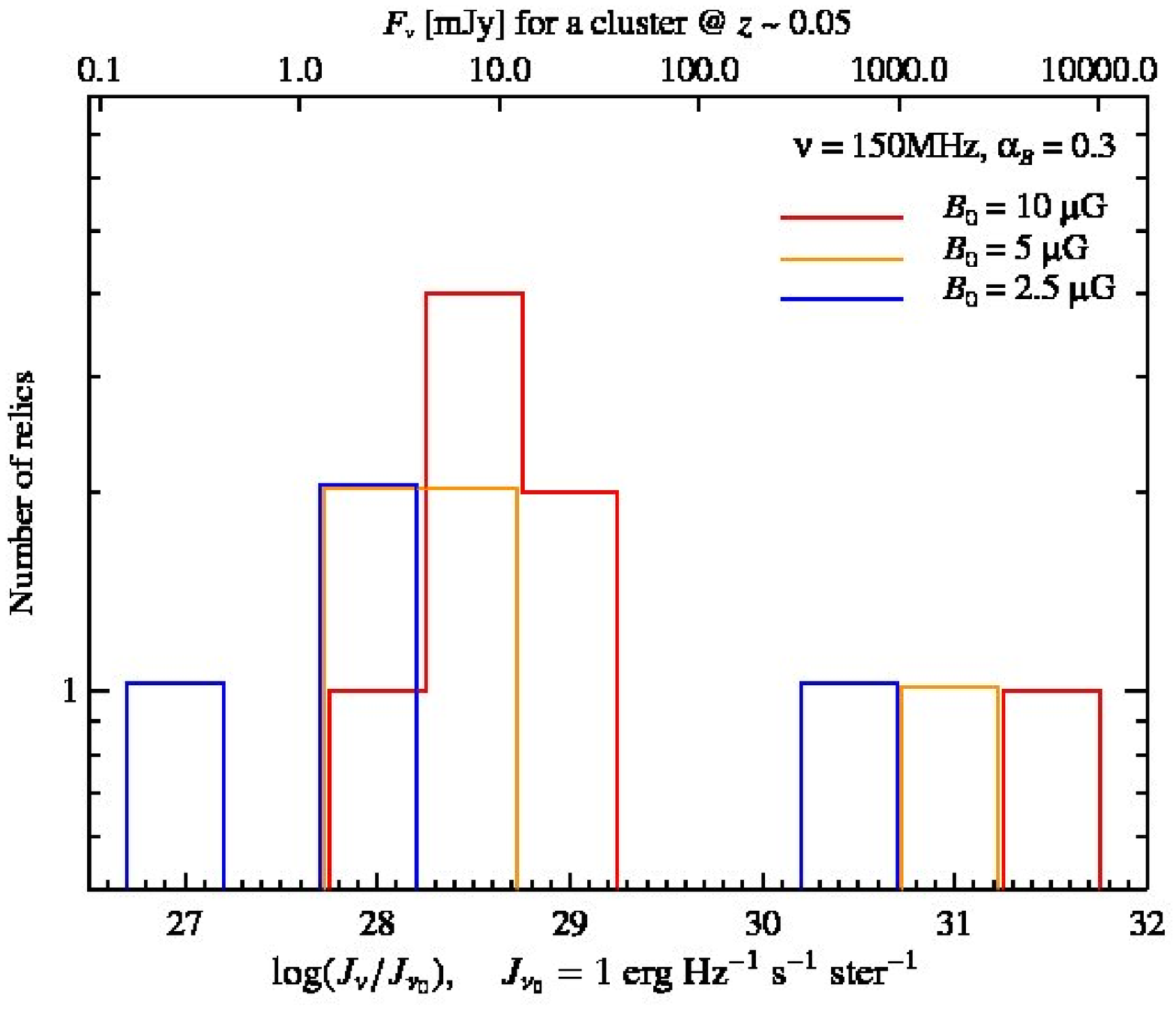}}%
\resizebox{0.5\hsize}{!}{\includegraphics{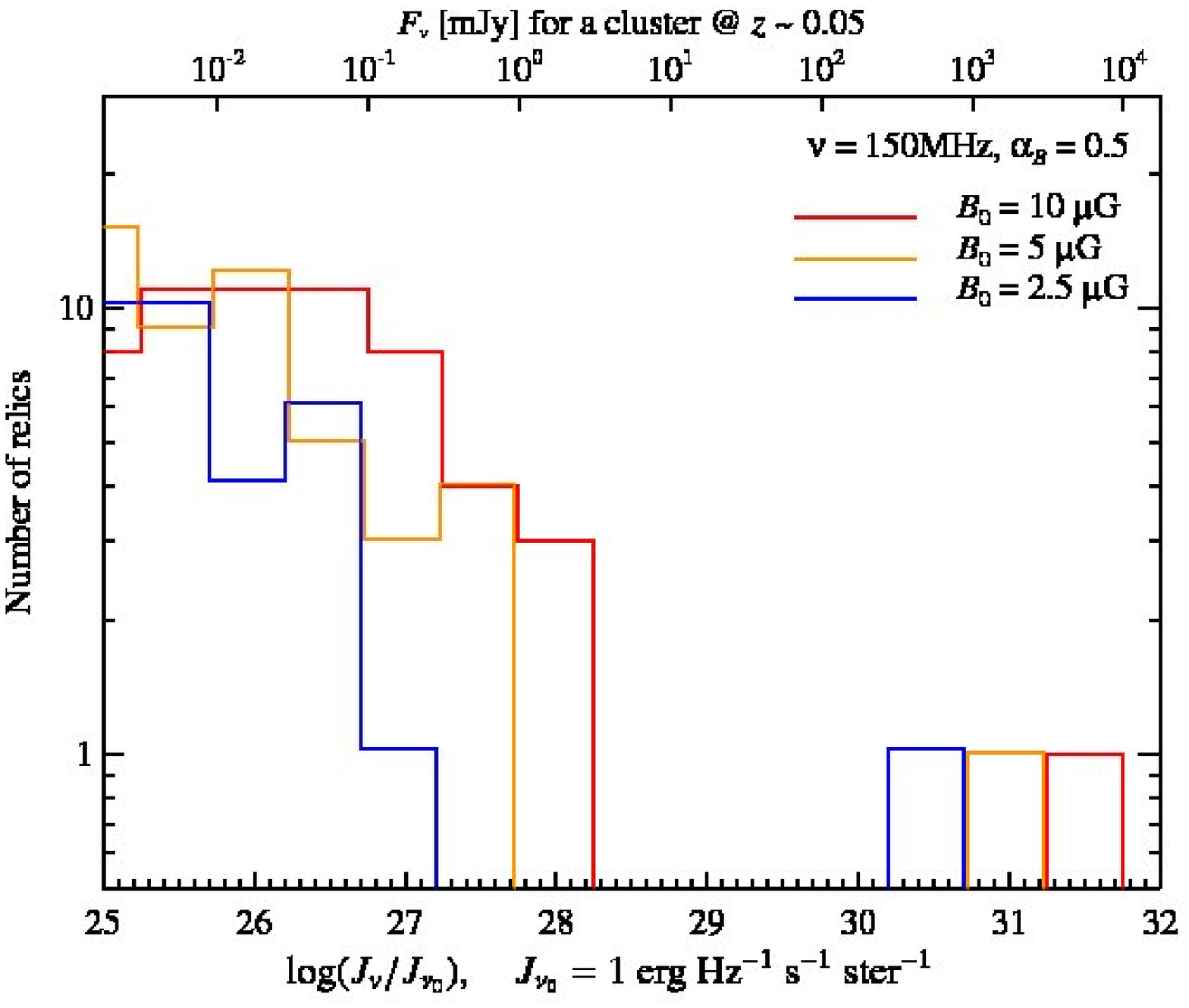}}\\
\end{center}
\caption{Luminosity functions with {\em observable parameters} (left) and {\em
    theoretical parameters} (right) of our relic finder.  The top and bottom
  panels show how the magnetic decline $\alpha_B$ and the magnetic core energy
  density $B_0$ impact the luminosity functions, respectively. The choice of
  magnetic field parameters has a large impact on the shapes of the
  luminosity function which is a consequence of the inhomogeneous nature of
  virializing processes in cosmic structure formation. Smaller
  $\alpha_{B}$ values, corresponding to a shallower magnetic decline,
  produce many more brighter relics compared to larger values of
  $\alpha_{B}$, which produce slightly more less-luminous
  relics. Increasing $B_0$ results in a greater number of more-luminous relics.
  Notice the loss of the less-luminous relics from the {\em observable} to
  {\em theoretical parameters}.  This is a result of the brightest relic
  swallowing up smaller relics due to a decrease of the emission threshold
  parameter.}
\label{fig:lfs_g72}
\end{figure*}

\subsubsection{Removal of galaxy contamination and cool core}
\label{sec:removal}

Our radiative simulations model star formation \citep{2003MNRAS.341.1253H}
which leads to the formation of galaxies. When applying the criteria described
in Sect.~\ref{sec:relicfinder}, these galaxies appear as false radio
relic candidates since they are dense, compact and have enough emissivity per
particle to be selected by the relic finder. To select against those objects,
we impose further constraints on the SPH particles that are grouped together
and require them to have a zero fraction of neutral hydrogen and to be below a
very conservative threshold of number density $n_\rmn{thres} = n_\rmn{SFT} / 32
= 0.004 \mbox{ cm}^{-3}$.

Our relic finder also picks up the over-cooled centres of galaxy
clusters in our simulations, contaminating the radio emission. Since the
candidate relic in the over-cooled center may be physically connected
to other true relics, 
it cannot be removed by simply discarding the closest relic candidate to the
center. We apply a very conservative cut in radius of $r = 40\mbox{ kpc}$ and
neglect the weak dependence on cluster mass and dynamical state. We note that
smaller clusters ($M<5\times10^{14}\rmn{M}_\odot$), in particular those with
dynamical activity, tend to have slightly smaller cooling regions.

\section{Results}
\label{sec:results}

\subsection{Probing the intra cluster magnetic fields}

In this section, we investigate how sensitive different radio synchrotron
observables are with respect to the properties of the large scale magnetic
field.

\subsubsection{Luminosity functions}

\begin{figure*}
\begin{center}
 \begin{minipage}[t]{0.495\textwidth}
   \centering{\it \large Rotation measure}
 \end{minipage}
 \hfill
 \begin{minipage}[t]{0.495\textwidth}
   \centering{\it \large Rotation angle}
 \end{minipage}
\resizebox{0.5\hsize}{!}{\includegraphics{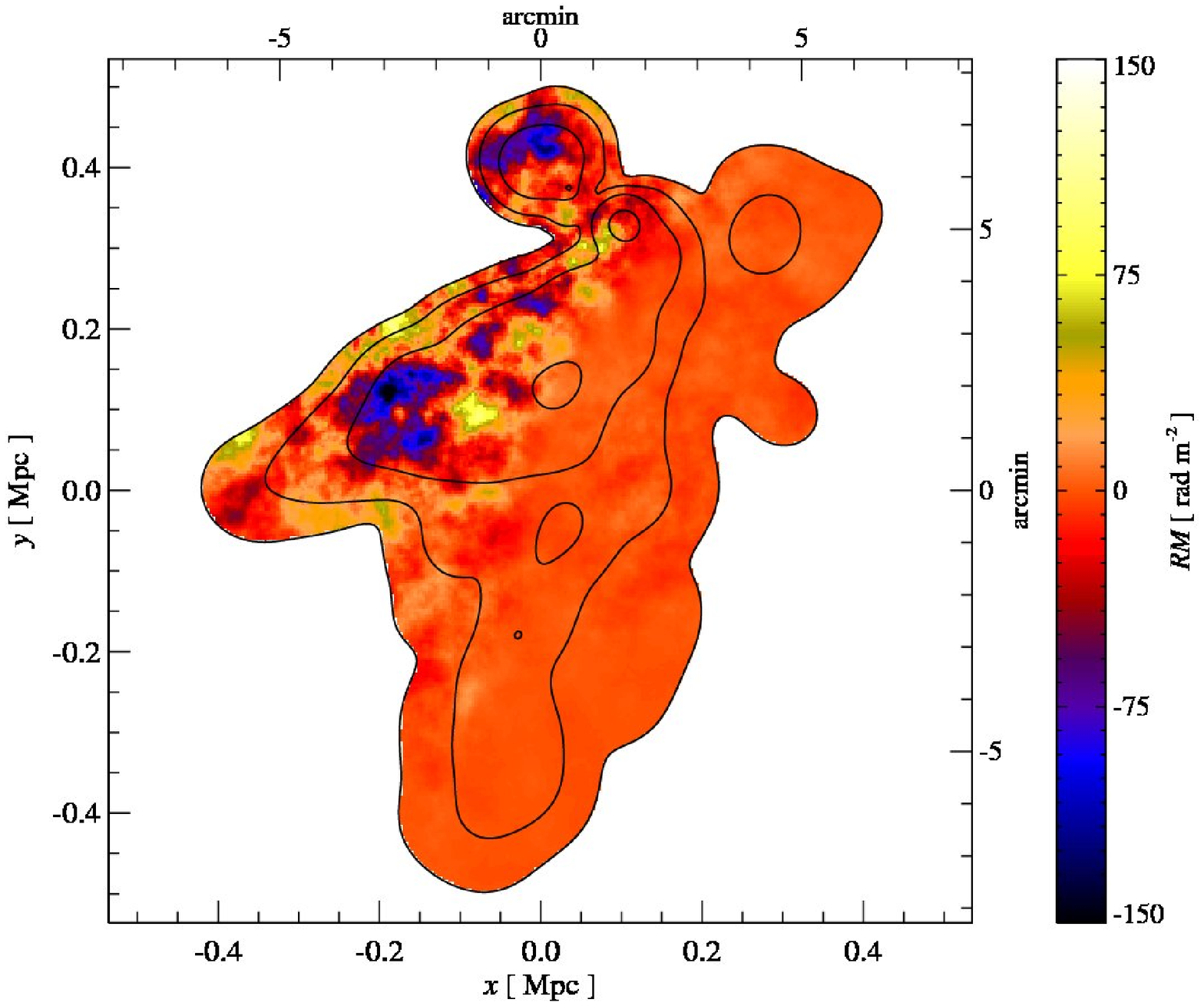}}%
\resizebox{0.5\hsize}{!}{\includegraphics{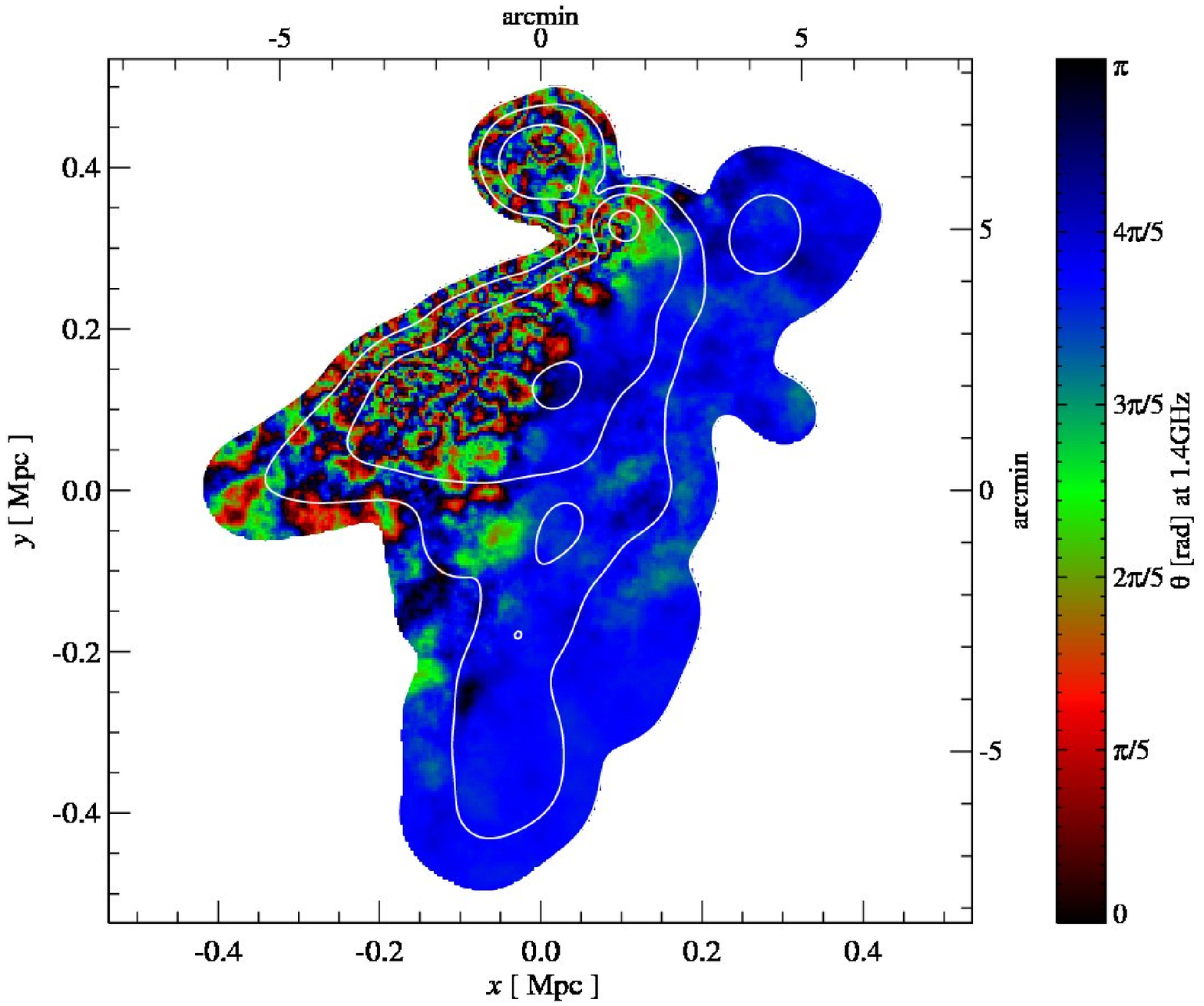}}\\
\end{center}
\caption{Left panel: Faraday rotation measure map of the largest relic
  in g72a, if the cluster were placed at z$\sim$0.05 (mimicking A2256). Right panel: Polarization
  angle map at $\nu = 1.4$ GHz, assuming a uniform rotation angle at the position
  of the relic. The cluster center is located in the direction of the upper left
  corner for both the images. Regions with high RM variance correspond to high
  spatial variation of the polarization angle. In combination with a finite beam
  size, this makes it challenging to observe a high degree of polarization in
  such a relic.  The magnetic field parameters are $\alpha_{B}$ = 0.7
  $B_0$ = 2.5 $\umu$G and the contours represent the surface brightness
  increasing in decades from $5\times 10^{-3}$ mJy arcmin$^{-2}$ at $\nu$ = 1.4 GHz.}
\label{fig:RM}
\end{figure*}

The magnetic fields within our simulations are parametrized by a
simple scaling relation.  For each cluster, we compute radio
luminosity functions to aid in differentiating between different
magnetic field parametrizations by employing the dependency of
synchrotron emissivity on the magnetic field of the ICM. Our
luminosity functions are distribution functions of the total
luminosity per relic ($J_{\nu}$), where
\begin{equation}
J_{\nu} = \sum_{a} j_{\nu,a} \frac{M_{a}}{\rho_{a}} = \sum_{a} J_{\nu,a}.
\label{eq:jnu_tot}
\end{equation}
The units of $J_{\nu,a}$ are erg s$^{-1}$ Hz$^{-1}$ ster$^{-1}$, $M_a$ and $\rho_a$ are the
SPH gas particle mass and density respectively for the set of SPH particles
within the relic labelled by $a$.

The number of relics seen depends on the magnetic field parametrization. In
Fig.~\ref{fig:lfs_g72}, we show how the luminosity functions depend on $B_0$
and $\alpha_{B}$. As expected, we find more and brighter radio relics
for higher values of $B_0$. However, rather than simply scaling the luminosity
function to higher relic emissivities for larger $B_0$ (assuming a fixed slope
$\alpha_{B}$), we find that their shapes change. This is a consequence
of the inhomogeneity of the virializing cosmic structure formation waves that
are illuminated by the synchrotron emitting electrons. The effect of varying
$B_0$ is analogous to the water level within a very inhomogeneous landscape
that corresponds to the strength of the virializing shock waves. This level can
adopt different values depending on the magnetic realization such that the
resulting synchrotron emitting objects end up single connected or disjoint. So,
one could consider using Minkowski functionals to characterize the different
relics.

The trend for $\alpha_{B}$ is the opposite: higher values of
$\alpha_{B}$ lead to a lower number of radio relics. The parameter
$\alpha_{B}$ represents the slope of the magnetic scaling
($\varepsilon_{\mathrm{th}_0} > \varepsilon_{\mathrm{th}}$), and a
steeper slope will result in the magnetic field strength falling off
faster with radius. We expect the effect of both $B_0$ and
$\alpha_{B}$ on the luminosity function to be a generic effect for all
the simulated clusters, since $j_{\nu} \propto B^{\alpha_{\nu} +
1}/(\varepsilon_B + \varepsilon_{\mathrm{CMB}})$
(cf. Eqns.~\ref{eq:f_eq} and \ref{eq:jnu}) and in the peripheral
cluster regions where $\varepsilon_B < \varepsilon_{\mathrm{CMB}}$, we
obtain $j_{\nu} \propto B^{\alpha_{\nu} + 1}$. The luminosity
functions alone are not sufficient to fully disentangle the magnetic
field properties, this will require other observables.

\subsubsection{Rotation measure}

Another independent approach to constrain magnetic field models are
Faraday rotation measurements.  Theoretically, one expects the
magnetic field in shocks to be aligned with them due to shock
compression \citep{1998A&A...332..395E} and stretching and shearing
motions induced by oblique shocks \citep{2006PhPl...13e6501S}. In
combination with the small synchrotron emitting volume that is caused
by the small synchrotron cooling time, this yields to polarized relic
emission.  Indeed, radio relics have been observed to be polarized up
to the 40 per cent level
\citep{2004NewAR..48.1137F,2006AN....327..553C}.
When polarized radio
emission propagates through a magnetized medium, its plane of
polarization rotates for a nonzero line-of-sight component of the
magnetic field $B_z$ due to the birefringent property of the
plasma -- Faraday rotation. The Faraday rotation angle is given by
\begin{equation}
\Phi_{\mathrm{obs}} = \lambda^2 RM + \Phi_{\mathrm{init}},
\label{eq:RMang}
\end{equation}
where
\begin{eqnarray}
\label{eq:RM}
  RM(\vecbf{x}_\bot) &=& a_0
  \int_0^L B_z(\vecbf{x})\, n_{\mathrm{e}}(\vecbf{x})\, \dd z\\ 
  \label{eq:RM2}
  &\simeq& 812\, \frac{\rmn{rad}}{\rmn{m}^2}\, \frac{B}{\umu G} \,
  \frac{n_{\mathrm{e}}}{10^{-3}\mbox{ cm}}\, \frac{L}{\mbox{Mpc}},
\end{eqnarray}
where $ a_0=e^3/(2\upi m_\e^2c^4)$, $\vecbf{x} = (\vecbf{x}_\bot,l)$, and
$n_{\mathrm{e}}$ is the number density of electrons. In Eqn.~\ref{eq:RM2}, we
have assumed constant values and a homogeneous magnetic field along the
line-of-sight to give an order of magnitude estimate for RM values. Assuming
statistically homogeneous and isotropic magnetic fields, the RM dispersion
$\bra RM^2\ket$ reads as follows,
\begin{eqnarray}
  \label{eq:RMsquared1}
  \bra RM^2 \ket &=& a_0^2 \left\bra \left[\int_0^L
  n_\e (\vecbf{x})  B_z (\vecbf{x})\, \dd z \right]^2 \right\ket \\
  \label{eq:RMsquared2}
   &=& a_0^2 \frac{3\lambda_B}{2} L \,\bra n_\e (\vecbf{x})^2  B_z (\vecbf{x})^2\ket\\
  \label{eq:RMsquared3}
   &=& a_0^2 \frac{3\lambda_B}{2} L\,\mathcal{C}_{n_{\mathrm{e}}B_z} \,
  \bra n_\e (\vecbf{x})^2\ket \bra B_z (\vecbf{x})^2\ket ,
\end{eqnarray}
\noindent where 
\begin{equation}
  \label{eq:cor}
  \mathcal{C}_{n_{\mathrm{e}}B_z} = 
  \frac{\bra B_z^2 n_{\mathrm{e}}^2 \ket}{\bra B_z^2 \ket \bra n_{\mathrm{e}}^2 \ket },
\end{equation}
\noindent is defined as the correlation factor and $\lambda_B =  2/3\times\lambda_z$ is the 3D
magnetic auto-correlation scale \citep{2003A&A...401..835E} which can be
estimated from the measured $RM$ power spectrum.

We studied the Faraday rotation of the largest relic in cluster g72a
(Fig.{\ }\ref{fig:RM}).  The aim is to recover intrinsic statistical
properties of the ICM magnetic field by studying RM statistics. We
produce the RM map by projecting the line-of-sight component of our
tangled magnetic field (Sect.~\ref{sec:mag-model}) and the thermal
electron density that has also been mapped from its Lagrangian
distribution onto a 3D grid.\footnote{Note that we neglect gas above
the very conservative threshold of number density $n_\rmn{thres} =
n_\rmn{SFT} / 32$ to be consistent with our proceeding in
Sect.~\ref{sec:removal}.}  Firstly, we are concerned about the
observability of polarized emission. The RM scales with $B_\parallel$
and $n_\e$ according to Eqn.~\ref{eq:RM} such that we expect RM values
to increase across the relic towards the projected cluster center.
However, large RM values leads to confusion when trying to observe the
polarization angle. Beam width depolarization
\citep{1966ARA&A...4..245G} takes place if the polarization angle
changes by a radian on scales shorter then the beam. To avoid this,
one can go to shorter wavelengths and smaller beams at the expense of
radio luminosity. Secondly, we are concerned with RM contamination
from the galaxy. Even at moderately high galactic latitudes the
galactic RM contribution \citep{1981ApJS...45...97S} can be
approximately the same order as the RM we calculate. However, the
strength of our RM depends on the relic location with respect to the
cluster and observer, as well as our magnetic field
parameters. Therefore, a different location or parametrization will
lead to stronger or weaker RM. Also, the galactic RM in principle can
be modelled and removed from the RM map.

\begin{figure}
\resizebox{\hsize}{!}{\includegraphics{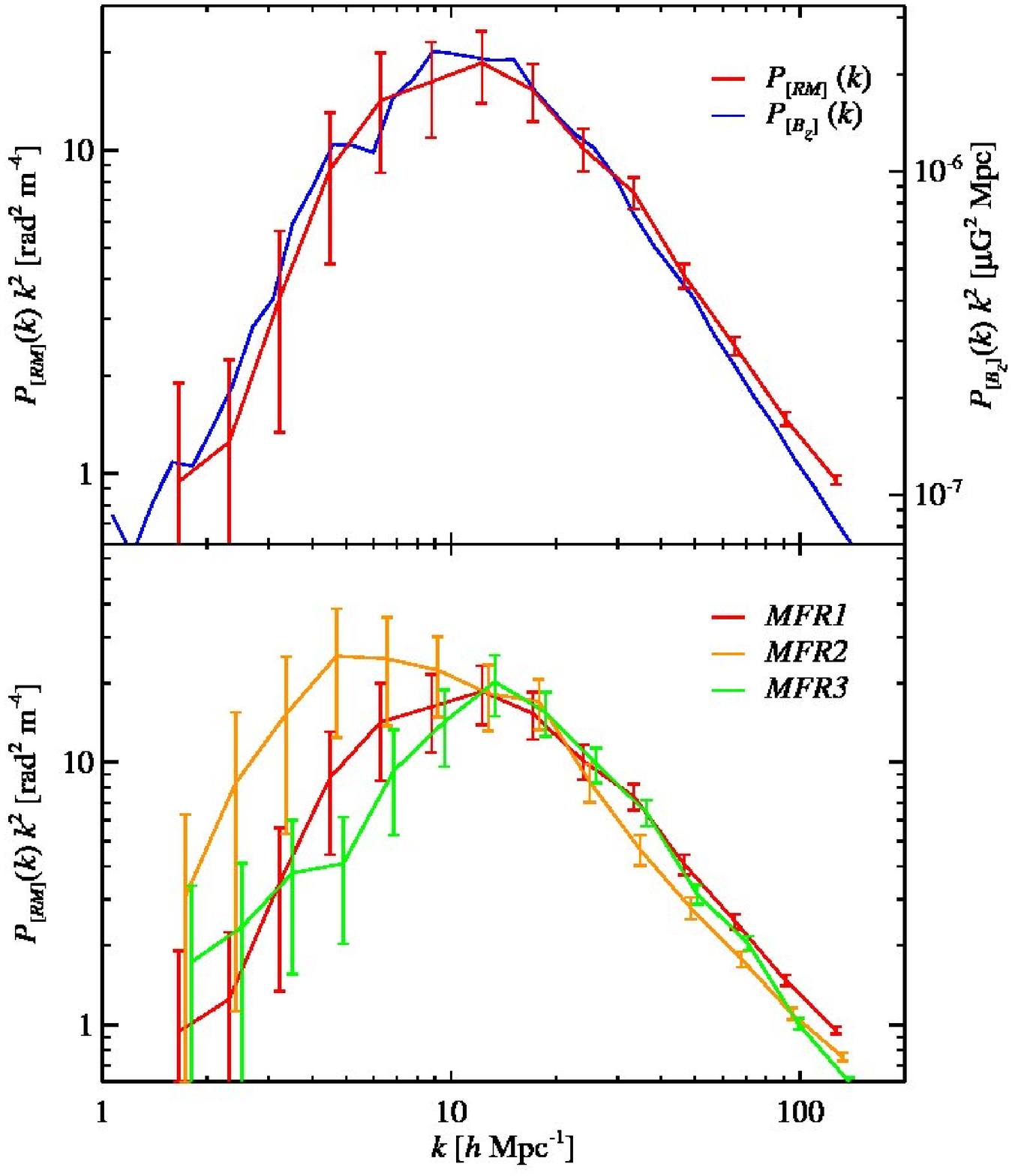}}
\caption{Top: power spectrum of the RM map $P_{[RM]}(k)$ and power spectrum of
  the line-of-sight component of the magnetic field $P_{[B_z]}(k)$ for MFR1,
  where the error bars represent the 1$\sigma$ confidence regions. $P_{[RM]}(k)$
  attains excess power at large angular scales from fluctuations in
  $n_{\mathrm{e}}$. Our RM power spectrum matches the shape and the peak scale
  of the input power spectrum within the error bars. Bottom: power
  spectra of RM maps for different magnetic field realizations
  (cf. Table~\ref{tab:bfield}). All RM power spectra recover the shape and
  characteristic scale of their magnetic input power spectra.}
\label{fig:powerspec}
\end{figure}

\begin{table}
\begin{center}
\caption{Magnetic field realizations (MFR) used in RM maps.}
\begin{tabular}{|c|c|c|}
\hline magnetic field & correlation length & input slope of \\ 
       realization name & ($\lambda_{B}$) [$h^{-1}$ kpc] & power spectra \\ 
\hline
 MFR1 & 100 & -5/3 \\ 
 MFR2 & 200 & -5/3 \\ 
 MFR3 & 100 & -2 \\ \hline
\label{tab:bfield}
\end{tabular}
\end{center}
\end{table}

High-quality rotation measure maps enable one to measure the RM power
spectra. The large angular extent of giant radio relics provides a powerful
tool of probing the maximum coherence scales of the magnetic field in clusters;
in contrast RM maps from radio lobes are typically much smaller.  We
calculate power spectra from our RM maps and magnetic field realizations for
each model separately (Fig.~\ref{fig:powerspec} and Table~\ref{tab:bfield}). For
consistency reasons, we only consider the volume subtended by the radio relic
when calculating the magnetic power spectrum.  We define the RM power spectra
($P_{[RM]}$) and the power spectrum of $B_z$ ($P_{[B_z]}$) as follows,
\begin{eqnarray}
  \label{eq:RMS1}
  \bra RM^2 \ket  &=& 2\pi \int_0^\infty k^{} P_{[RM]}\, \dd k , \\
  \label{eq:RMS2}
  \bra B^2 \ket &=& 3\cdot 4\pi \int_0^\infty k^2 P_{[B_z]} \, \dd k,
\end{eqnarray}
where the additional factor of three accounts for fluctuations in the total
magnetic field while assuming our random Gaussian field.  A partial Monte Carlo
method was used to determine the 1$\sigma$ error bars on the power
spectrum. Assuming a constant magnetic coherence scale, we construct the
envelope function by computing the variance of RM (Eqn. \ref{eq:RMsquared2}).  We
multiply this envelope function with $10^3$ realizations of random Gaussian
field and measure the power spectrum on each of these maps.  The fractional
errors are computed from the variance of these power spectra. While the first
six power spectrum bins are on average 50 per cent correlated, the correlations
drop to be below 20 per cent for the bins on smaller scales.

By construction our parametrization of the magnetic field is
correlated with the electron number density $n_{\mathrm{e}}$ which
might possibly introduce biases in our RM maps. However, comparing RM
power spectrum to the magnetic power spectrum, we find that the RM
power spectrum overall resembles the original shape of the magnetic
power spectrum and the injection scale corresponds to the scale of
maximal power in the RM map. This is true for all our magnetic field
realizations. We measure the slope of these power-laws at small scales
to an accuracy of $\pm$0.05, which is significant to differentiate
between Kolmogorov ($k^{-5/3}$) and Burgers ($k^{-2}$) turbulence
spectra. We also find that the measurement of the RM power spectrum
slope is independent of the magnetic field parameters $\alpha_{B}$ and
$B_0$. The full correlation matrix of the power spectrum bins is used
when fitting the power-laws and in the error calculations. We find
that our measured slope is flatter then expected and can be attributed
to small scale fluctuations in $n_{\mathrm{e}}$, since $P_{[B_z]}$ has
the same slope as the power spectrum of $B_z$ integrated along the
line of sight. The small inhomogeneity of $n_{\mathrm{e}}$ in our
simulations does not severely affect the intrinsic spectral
shape. Thus, it is possible in principle to recover the intrinsic 3D
magnetic power spectrum by solving the inverse problem \citep{
2003A&A...412..373V, 2003A&A...401..835E, 2005A&A...434...67V}.

Using Eqns. \ref{eq:RMsquared3}, \ref{eq:RMS1} and \ref{eq:RMS2} we
estimated the {\em rms} magnetic field strength from $P_{[RM]}$ and $P_{[B_z]}$
respectively and recover our initial {\em rms} magnetic field strength. We find
that our correlation factor ($\sqrt{C_{n_{\mathrm{e}}B_z}} \simeq
5.3$) \footnote{We caution the reader that the particular value of
  $C_{n_{\mathrm{e}}B_z}$ reflects the parametrization of the magnetic
  field we adopt in our model, and may be realized differently in
  Nature. We also note that the value 
  of the correlation factor in the non-radiative simulation by
  \citet{2008MNRAS.tmp..277P} is $\sqrt{C_{n_{\mathrm{e}}B_z}}\simeq
  6$. Further work is required to address this question in the context of MHD
  cluster simulations.} is 20 per cent larger than the correlation factor
obtained by fitting a smooth $\beta$-model to the spherically averaged profile
of $n_{\mathrm{e}}$ and scaling $B \propto n_{\mathrm{e}}^{\alpha_B}$ with the
same $\alpha_B=0.7$ that we used to construct our RM maps (similar to the
procedure applied by \citealt{2003A&A...401..835E} and
\citealt{2004A&A...424..429M}).  This result suggests that the fairly
homogeneous density distribution in our simulations (after removing the galaxy
contamination described in Sect.~\ref{sec:removal}) does not severely
bias the average magnetic field strengths estimated by RM studies if one takes
into account the overall shape of the profiles of $n_\e$ and $B$.

For convenience, we derive a formula for the {\em rms} magnetic field
strength ($\sqrt{\bra B^2 \ket}$) as a function of the peak of
$P_{[RM]}$ and the {\em rms} fluctuations of $n_\e$. Multiplying
$P_{[RM]}$ with a Heaviside function $\Theta(k)$ and ensuring that the
intrinsic spectrum is sufficiently steeper than $k^{-2}$,
Eqn.~\ref{eq:RMS1} becomes
\begin{eqnarray}
  \label{eq:scaleB}
  \bra RM^2 \ket  &\simeq& 2\pi \int_0^\infty k^{} P_{[RM]}(k)\,
  \Theta(k-k_\mathrm{peak}) \, \dd k ,\nonumber \\
  &\simeq& \pi P_{[RM]}(k_\mathrm{peak})\, k_\mathrm{peak}^{2}.
\end{eqnarray}

\noindent The value of $P_{[RM]}(k_\mathrm{peak})\, k_\mathrm{peak}^{2}$ can be
read off directly from Fig.~\ref{fig:powerspec} and combining
Eqns.~\ref{eq:scaleB} and \ref{eq:RMsquared3} yields an approximate value for
$\sqrt{\bra B^2 \ket}$,
\begin{eqnarray}
  \label{eq:scaleB2}
  \sqrt{\bra B^2 \ket} &\simeq& \sqrt{\frac{{2 \pi\,
  P_{[RM]}(k_\mathrm{peak})\, k_\mathrm{peak}^{2}}}{a_0^2
  \,\lambda_B\, L\,\mathcal{C}_{n_{\mathrm{e}}B_z} \, \bra
  n_\e^2\ket}}\\ &\simeq& 0.1\mu\mathrm{G}
  \left(\frac{P_{[RM]}(k_\mathrm{peak})\,
  k_\mathrm{peak}^{2}}{50\,\mathrm{rad}\,\mathrm{m}^{-2}}\right)^{\frac{1}{2}}
  \left(\frac{\sqrt{\bra n_{\mathrm{e}}^2 \ket}}{10^{-4}\,h^{2}\,
  \mathrm{cm}^{-3}}\right)^{-1}\nonumber\\ 
  &\times&
  \left(\frac{\lambda_{B}}{100\,h^{-1}\,
  \mathrm{kpc}}\right)^{-\frac{1}{2}}
  \left(\frac{L}{4\,h^{-1}\,\mathrm{Mpc}}\right)^{-\frac{1}{2}}
  \left(\frac{\mathcal{C}_{n_{\mathrm{e}}B_z}}{27}\right)^{-\frac{1}{2}},
\end{eqnarray}
where we inserted numerical values from our simulation in the last step.

\subsection{Existence and properties of the WHIM}

In this section, we investigate the potential of radio relic observations to
infer the hydrodynamic properties such as density and temperature of the WHIM.

\subsubsection{Properties of virializing shocks}
\label{sect:shocks}
\begin{figure*}
\begin{center}
 \begin{minipage}[t]{0.495\textwidth}
   \centering{\it \large Spectral index projection of largest relic}
 \end{minipage}
 \hfill
 \begin{minipage}[t]{0.495\textwidth}
   \centering{\it \large Observed spectral index projection}
 \end{minipage}
\resizebox{0.5\hsize}{!}{\includegraphics{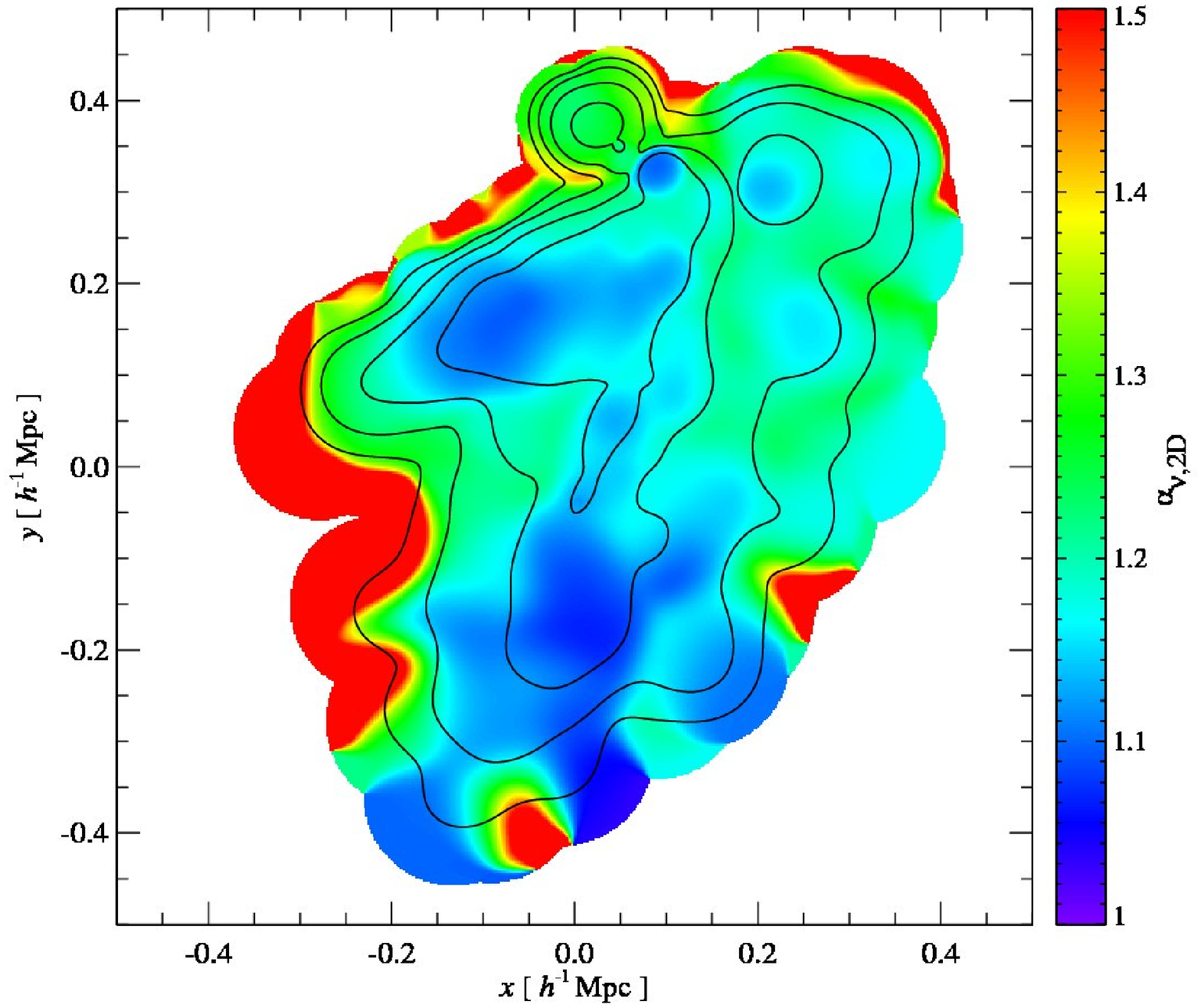}}%
\resizebox{0.5\hsize}{!}{\includegraphics{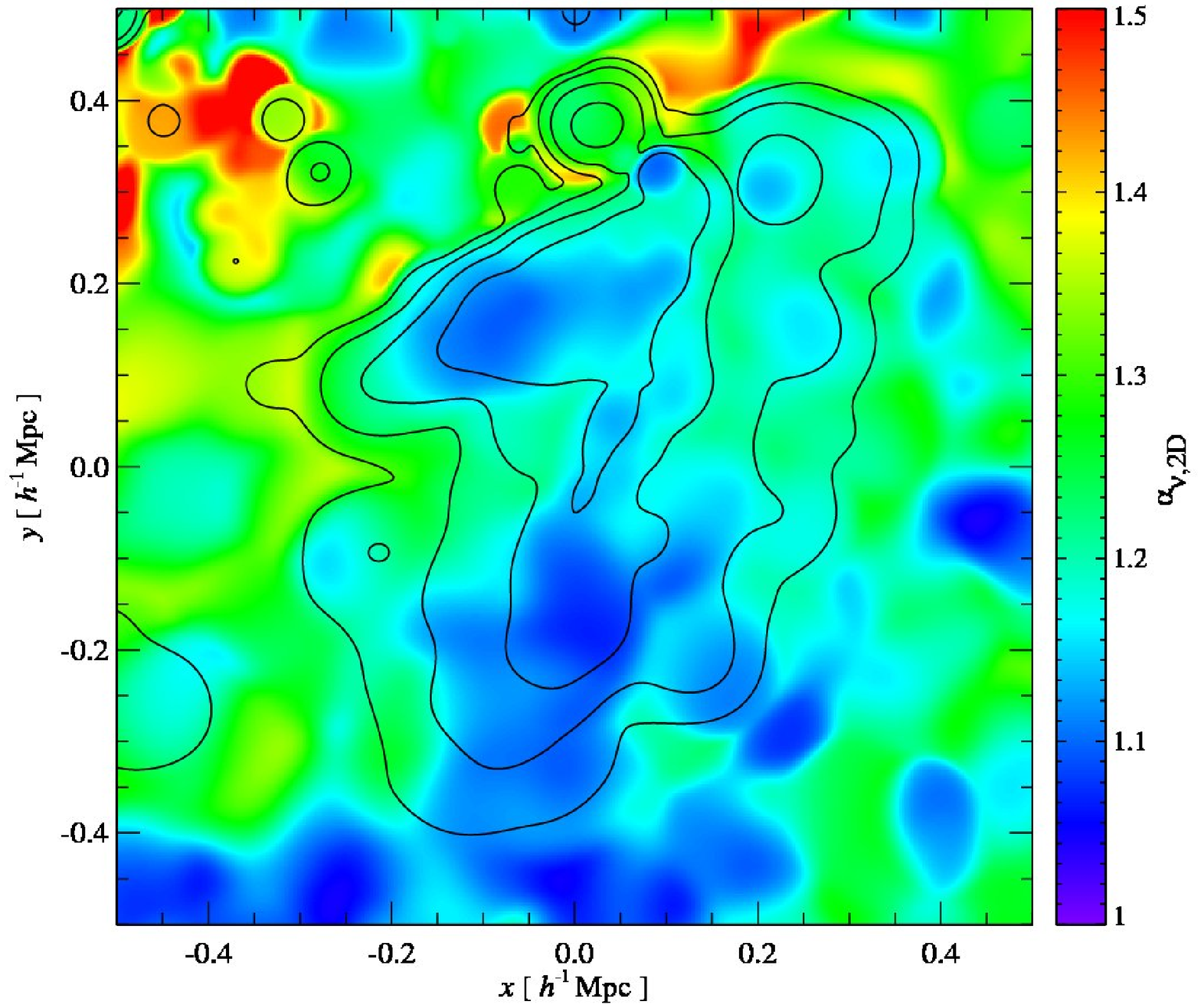}}\\
\end{center}
\caption{Spectral index map, $\alpha_{\nu,2D}$, between 150 MHz and 1.4 GHz for
  the largest relic in g72a, with only the SPH particles selected by the relic finder
  projected (left panel) and for the total emission in the same region (right
  panel). The contours show orders of magnitude in surface brightness
  in mJy arcmin$^{-2}$, with the
  highest contour representing 5 mJy arcmin$^{-2}$ at $\nu$ = 1.4 GHz. Notice the
  edge effects that show up in the projection of the single relic where the
  emission falls off. These effects are due to the sharp emissivity cutoff of
  our relic finder and incomplete sampling of SPH relic particles at different
  frequencies. More importantly, in regions with high synchrotron brightness,
  the spectral index is almost uniform across the central relic implying that
  this relic traces a single formation shock wave. }
\label{fig:proj}
\end{figure*}

\begin{figure*}
\begin{center}
 \begin{minipage}[t]{0.495\textwidth}
   \centering{\it \large 2D spectral index distributions}
 \end{minipage}
 \hfill
 \begin{minipage}[t]{0.495\textwidth}
   \centering{\it \large 3D spectral index distributions}
 \end{minipage}
\resizebox{0.5\hsize}{!}{\includegraphics{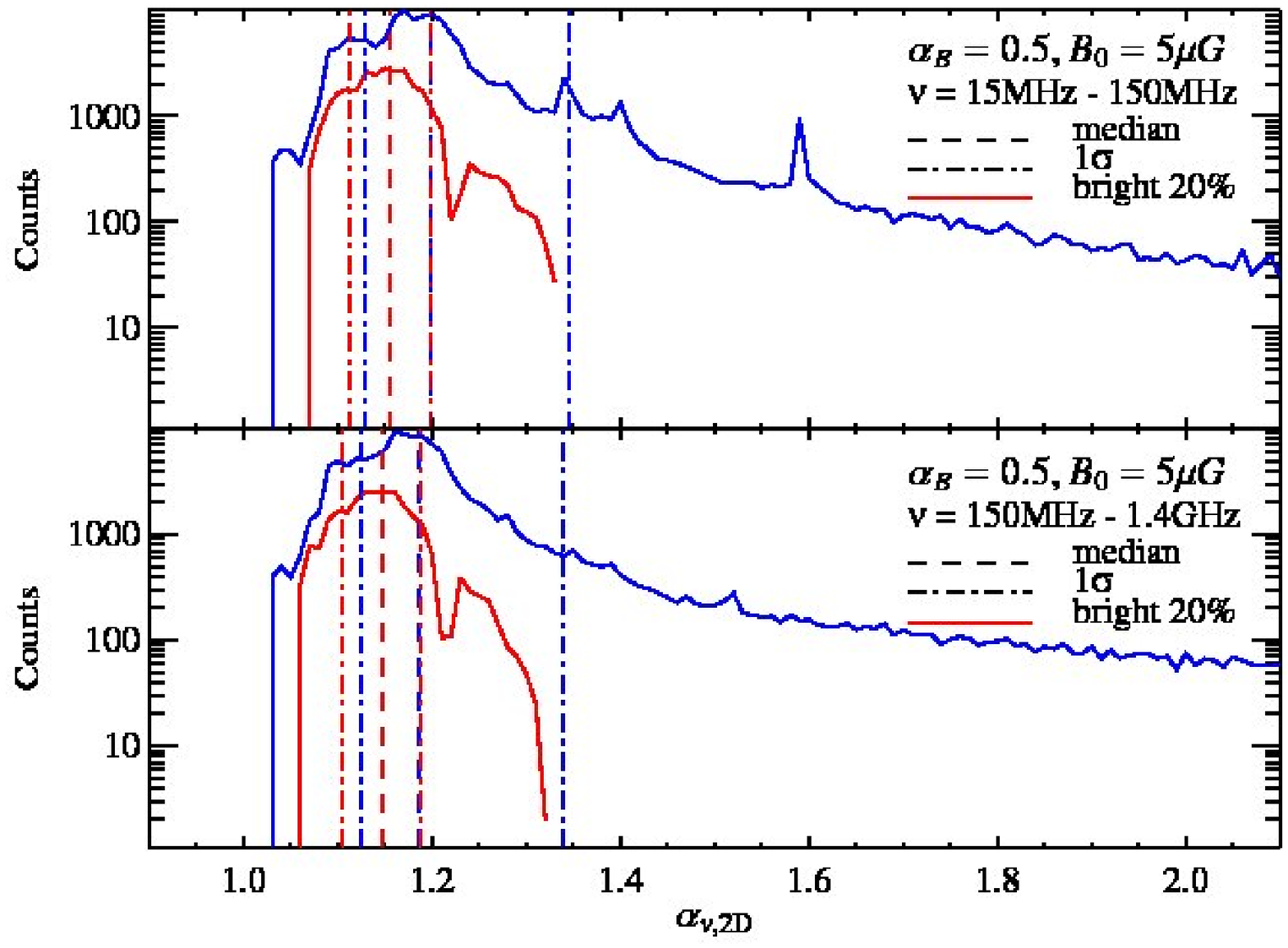}}%
\resizebox{0.5\hsize}{!}{\includegraphics{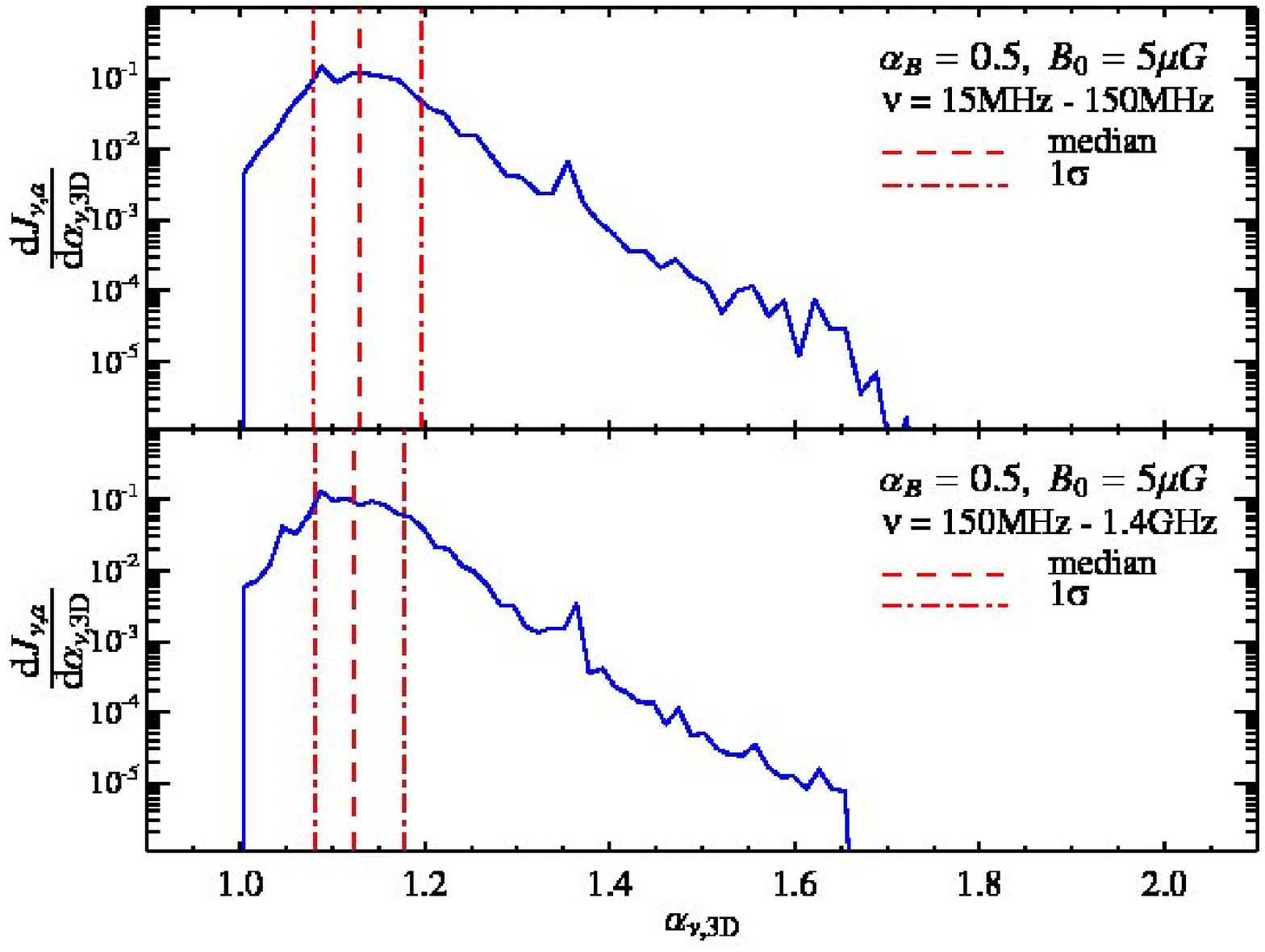}}\\
\resizebox{0.5\hsize}{!}{\includegraphics{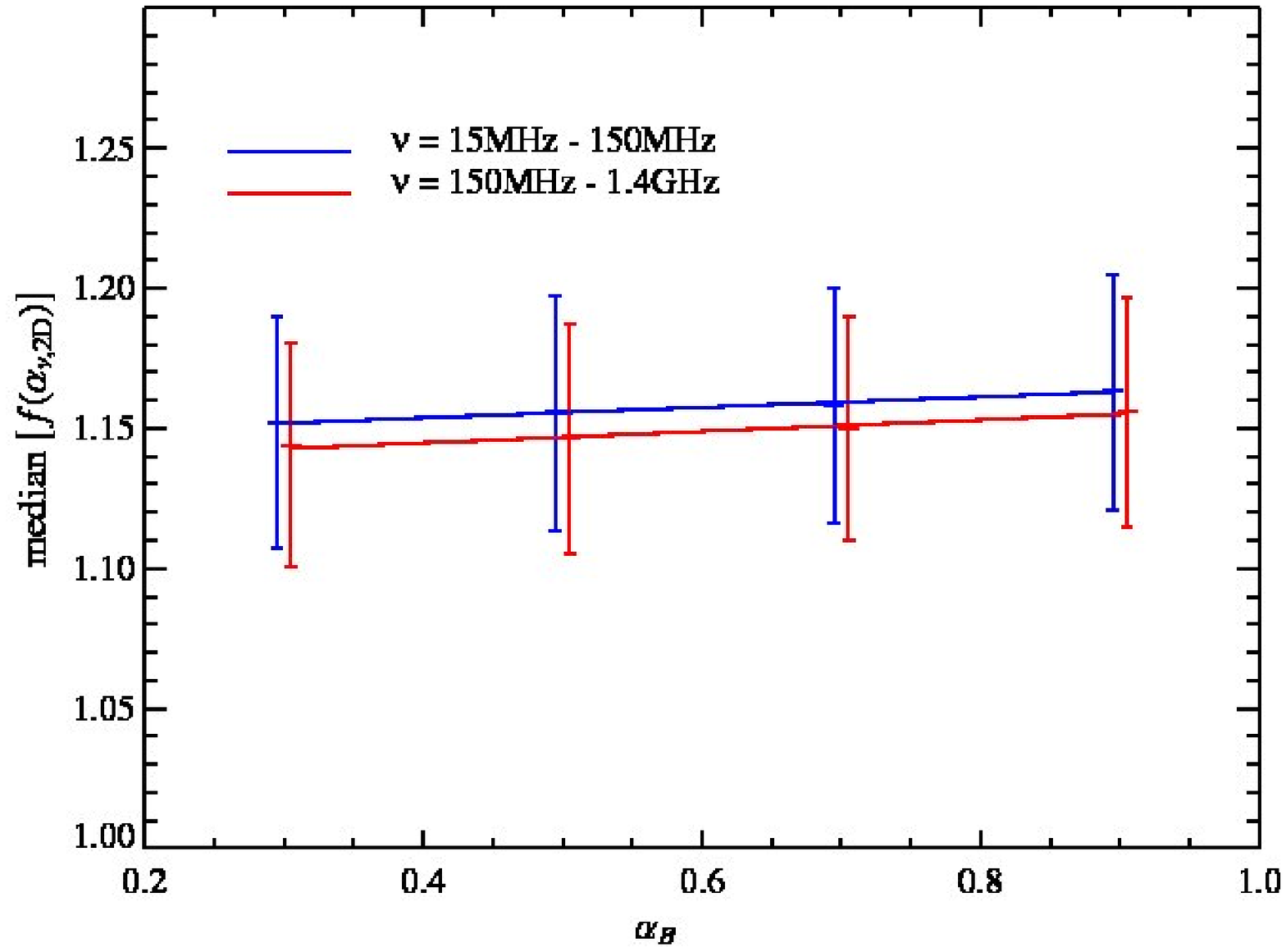}}%
\resizebox{0.5\hsize}{!}{\includegraphics{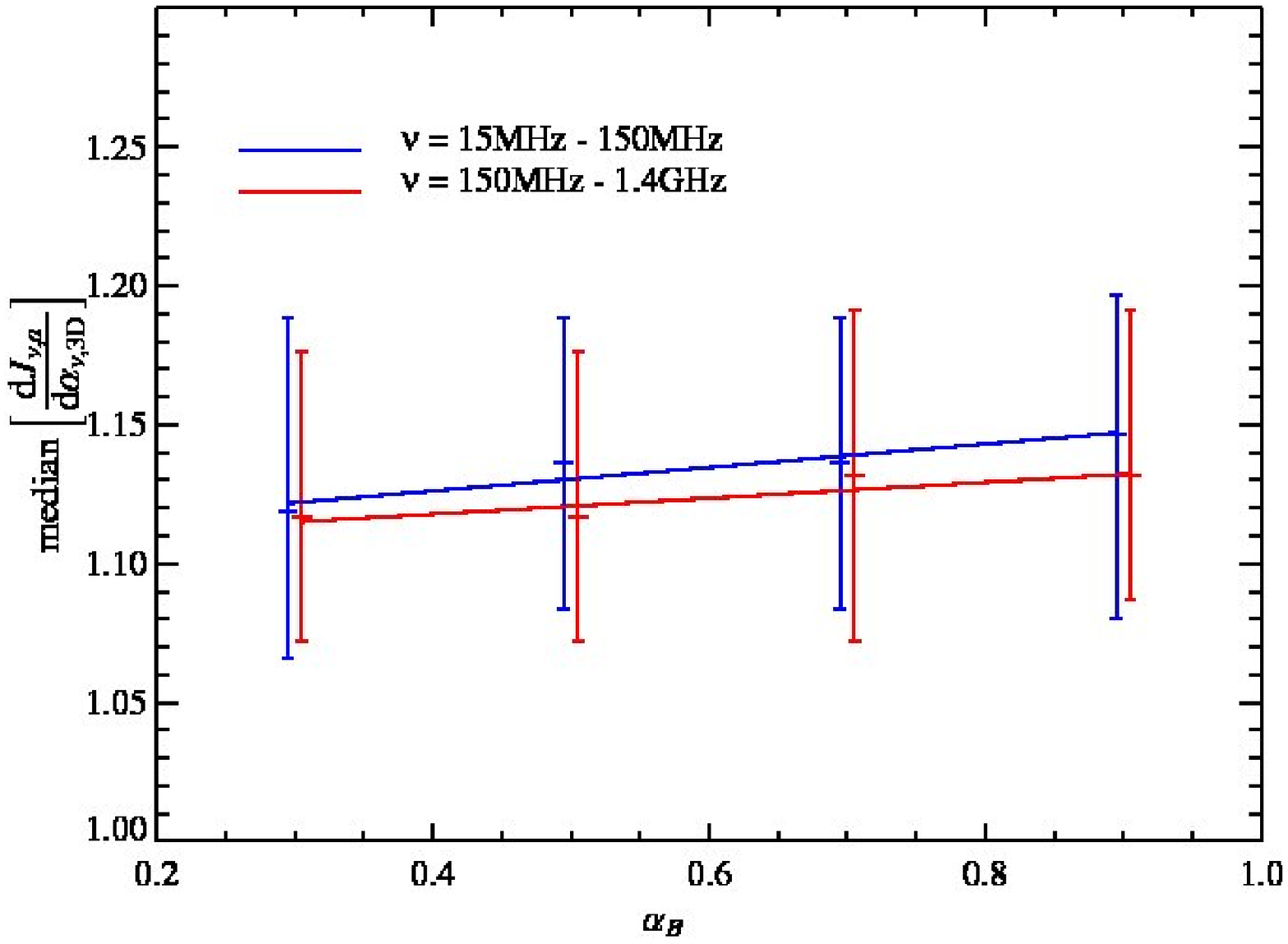}}\\
\end{center}
\caption{Top: spectral index distributions for our largest relic using our
  standard particular magnetic field parametrization. 2D spectral index
  distributions, $f(\alpha_{\nu\rmn{2D}})$, of the full map and the brightest 20
  percent pixels (top left) are contrasted to the radio luminosity weighted
  distribution of the 3D spectral index, $\dd J_{\nu,a} / \dd
  \alpha_{\nu,\rmn{3D}}$ (top right).  Bottom left: the median of
  $f(\alpha_{\nu\rmn{2D}})$ for the brightest 20 percent pixels ($S_\nu\geq$ 12
  mJy arcmin$^{-2}$ at $\nu$ = 150 MHz and $S_\nu\geq$ 1 mJy arcmin$^{-2}$ at $\nu$ =
  1.4 GHz) as a function of the magnetic decline $\alpha_B$ with the error bars
  representing the 1$\sigma$ percentiles. Bottom right: the median of $\dd
  J_{\nu,a} / \dd \alpha_{\nu,\rmn{3D}}$ as a function of the magnetic
  decline $\alpha_B$ with the error bars representing the 1$\sigma$
  percentiles.  This shows that for the giant radio relic of g72a, the median of
  the 2D and the 3D spectral indices agree statistically and are almost
  independent of the magnetic decline $\alpha_B$.}
 \label{fig:alphas3}
\end{figure*}

Diffusive shock acceleration determines the shape of the CRe spectrum that we
model as a power-law momentum spectrum (neglecting non-linear
effects). Synchrotron losses cause a steepening of this power-law by one power
of momentum. Spatially inhomogeneous virializing shocks with a distribution of
shock strengths cause a spatial variation of the spectral index of the cooled CR
electron spectrum. This is reflected in an inhomogeneous distribution of
synchrotron spectral index that may help to reconstruct the merging geometry by
providing a snapshot of the structure formation process in a galaxy cluster.

The spectral indices of the radio surface brightness $\alpha_{\nu,\mathrm{2D}}$
and that of the intrinsic 3D emissivity $\alpha_{\nu,\mathrm{3D}}$ are defined by,
\begin{equation}
\alpha_{\nu,\mathrm{2D}} = 
-\frac{\log(\frac{S_\nu}{S_{\nu_\mathrm{o}}})}{\log(\frac{\nu}{\nu_\mathrm{o}})},
\label{eq:alpha2}
\end{equation}
\begin{equation}
\alpha_{\nu,\mathrm{3D}} = 
-\frac{\log(\frac{J_{\nu,a}}{J_{\nu_\mathrm{o},a}})}{\log(\frac{\nu}{\nu_\mathrm{o}})}.
\label{eq:alpha3}
\end{equation}
It is unclear ab initio whether the projected spectral index
represents the actual deprojected quantity
($\alpha_{\nu,\mathrm{3D}}$) due to possible superposition of
different radio emitting structures along the line-of-sight.  We study
how these two quantities relate to each other, and present 2D spectral
index maps of both the largest radio relic in g72a and the total
emission from the same area (cf.{\ }Fig.{\ }\ref{fig:proj}). We note
that more than 99 per cent of the total radio emission can be
attributed to emission within the radio relic.  As a result, the
$\alpha_{\nu,\mathrm{2D}}$ maps are not contaminated by the diffuse
spurious emission.  Note the edge effects in Fig.{\ }\ref{fig:proj}
that show up in the projection of the single relic where the emission
falls off. These effects are due to the sharp emissivity cutoff of our
relic finder and incomplete sampling of SPH relic particles at
different frequencies.  In regions with high synchrotron brightness,
one can ignore these edge effects, and the resulting distribution of
$\alpha_{\nu,\mathrm{2D}}$ is fairly uniform
($\bra\alpha_{\nu,\mathrm{2D}}\ket \simeq$ 1.15, with $\sigma_{\alpha_{\nu,
\mathrm{2D}}} \simeq$ 0.04) implying that this relic traces a single
structure formation shock wave.

We further study the distribution of $\alpha_{\nu,\mathrm{2D}}$ and
$\alpha_{\nu,\mathrm{3D}}$ in our largest individual relic to uncover a
connection between them. Probability distribution functions (PDF) are
constructed for both $\alpha_{\nu,\mathrm{2D}}$ and $\alpha_{\nu,\mathrm{3D}}$
for varying parameters of the magnetic field (cf. Fig. \ref{fig:alphas3}). To
avoid contamination from edge effects seen in Fig. \ref{fig:proj}, the
$\alpha_{\nu,\mathrm{2D}}$ PDF was made for the brightest 20 per cent of the
pixels and the $\alpha_{\nu,\mathrm{3D}}$ PDF was weighted by particle
emissivity. These distributions do not change with our choices for magnetic
field parameters implying that the spectral indices are practically independent
of the magnetic field and depends mainly on properties of the shock. Another
striking result is that the median values for $\alpha_{\nu,\mathrm{2D}}$ and
$\alpha_{\nu,\mathrm{3D}}$ are statistically consistent within 1-$\sigma$.
Assuming that the line-of-sight integral is dominated by one bright relic and
choosing a pixel scale that is smaller than the length scale on which the
post-shock density varies, we can easily show that the 2D and intrinsic 3D
spectral index are identical. If there are more radio emitting regions
contributing to the observed surface brightness, we expect a concave radio
spectrum. Synchrotron cooling as well as re-acceleration lead to spectral
steepening in particular at high radio frequencies
\citep{1987A&A...182...21S}. Future work is required to address the associated
biases of the relation between $\alpha_{\nu,\mathrm{2D}}$ and
$\alpha_{\nu,\mathrm{3D}}$.

\begin{figure}
\resizebox{\hsize}{!}{\includegraphics{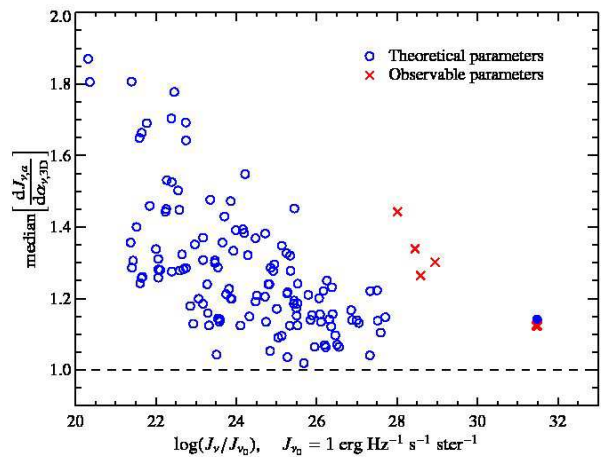}}
\caption{Two dimensional observable parameter space for radio relics in galaxy
  cluster g72a. Each symbol represents a relic within g72a and they are
  characterized by total luminosity and median of the 3D spectral index. Shown
  are {\em observable} relics (red crosses), {\em theoretical} relics (blue
  circles) and the large relic is emphasized by the bold cross and the filled
  circle. There is a trend to high spectral indices and a larger scatter for
  less luminous relics. As shown in Fig.~\ref{fig:lfs_g72}, the loss of the
  less luminous relics from the {\em observable} to {\em theoretical
    parameters} is a result of the brightest relic swallowing up these smaller
  relics due to a decrease of the emission threshold parameter.}
\label{fig:paramspace}
\end{figure}

Figure \ref{fig:paramspace} shows the observable parameters space of relic
luminosity and the median 3D spectral index. This parameter space compares the
shocks strength, which is related to the 3D spectral index (cf.{\
}Eqn.~\ref{eq:mach}) to the energy dissipated at the shock, which is related to
the relic luminosity. There is a trend that strong shocks are associated with
the more luminous relics. The implications of this trend is that the brightest
radio relics should show predominantly flatter spectral indices, which is the
current observational status of giant radio relics \citep{2008SSRv..134...93F}.

\subsubsection{Predicting pre-shock properties}

A majority of the hot gas ($> 10^7$ K) found at the centers of galaxy clusters
is believed to originate from the WHIM that is shock heated through large-scale
virializing structure formation shocks. These structure formation shocks are
traced by synchrotron emission in form of radio relics from recently accelerated
electrons (cf. Sect. \ref{sect:CRsync}). We have shown that under particular
conditions, the observed median 2D spectral index corresponds to the weighted
median 3D spectral index $\alpha_{\nu,\mathrm{3D}}$
(cf. Sect.~\ref{sect:shocks}).  The 3D spectral index can be related to the Mach
number of the shock (cf. Eqn.~\ref{eq:mach}) under the assumption that we have
an ideal fluid with a given adiabatic index. One can obtain information on the
post-shock values for density, pressure and temperature of the ICM through
deprojections of deep X-ray or Sunyaev-Zel'dovich observations
\citep{1998ApJ...500L..87Z}. With the knowledge of the Mach numbers combined
with post-shock values we calculate the pre-shock conditions of the ICM (the
WHIM) using the Rankine-Hugoniot jump conditions (cf. Appendix
\ref{sect:RHjump}).

\begin{figure}
\resizebox{\hsize}{!}{\includegraphics{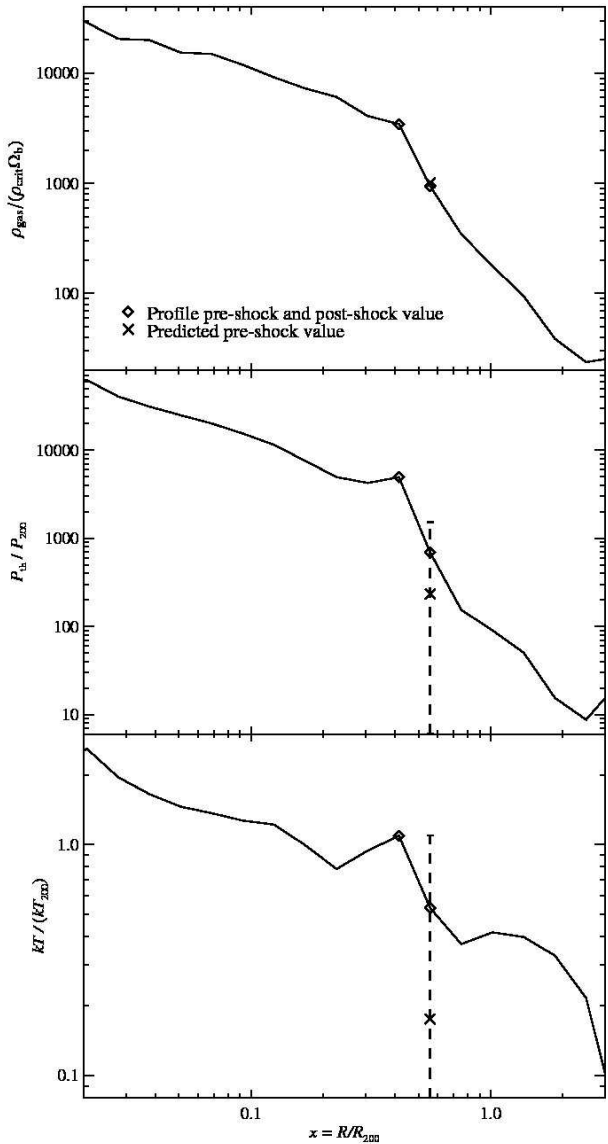}}
\caption{Radial profiles of galaxy cluster g72a restricted to the solid angle
  subtended by the largest relic for the density, pressure and
  temperature. The shocked region is seen in the profile at roughly $R_{200}/2$
  and is marked by the black diamonds with one sigma error bars, except for the
  density profile where the error bars are too small to show. The predicted
  pre-shock values (crosses) are $\sim$ 7\%, $\sim$ 67\% and $\sim$ 67\%
  different from the average profile values, but fall within the standard
  deviation for pressure and temperature. Most of this variation is caused by the
  shock being oblique and not perfectly tangential.}
\label{fig:WHIM}
\end{figure}

We take an optimistic approach and assume that the deprojections of
the thermal observables can be done ideally such that we use our
radial profiles calculated from the solid angle subtended by the
largest relic for simplicity (cf. Fig. \ref{fig:WHIM}). We define the
shock region by locating the radial bins that contain the majority of
shocked relic particles ($>85\%$). A small fraction of the relic
particles leak into radial bins adjacent to the shock causing slight
enhanced values of the radial profile. As mentioned above we insert
the calculated average Mach number and the post-shock value into the
Rankine-Hugoniot jump conditions for density, pressure and temperature
to estimate the upper limits of these WHIM properties. Our
predicted upper limits for pressure and temperature of the WHIM are
consistent with the simulated pre-shock properties within one standard
deviation. We note that the standard deviation of these hydrodynamic
properties reflect actual physical variations due to an oblique shock
that is not perfect tangential. Additionally, the particular relic
chosen is located at $\sim R_{200}/2$, which is within the cluster
volume. There are other observationally know relics that reside at the
virial radius and beyond
\citep[{\it e.g.} ][]{2006Sci...314..791B}. These relic are better
suited to probe the WHIM in combination with future X-ray and
multi-frequency SZ data. Thus, this example is to be taken as a
demonstration of our concept.

\begin{figure}
\resizebox{\hsize}{!}{\includegraphics{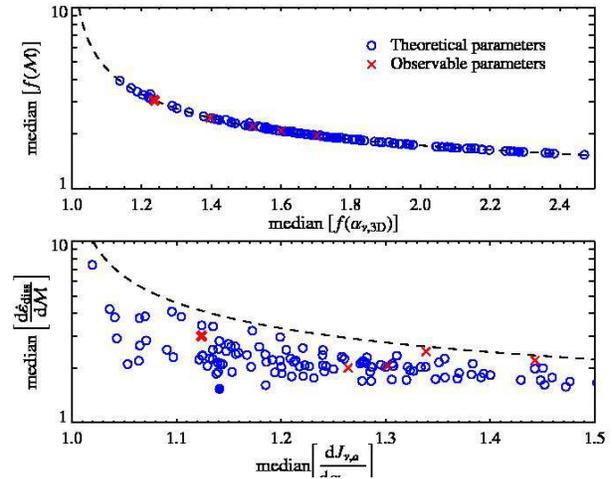}}
\caption{Top: median of the Mach number distribution for each relic as
  a function of the median of the distribution function of
  $\alpha_{\nu,\mathrm{3D}}$. Bottom: the observationally relevant
  quantities are the weighted distribution functions. Shown are median
  of $\dd \dot{\eps}_\rmn{diss} / \dd \mathcal{M}$ as a function of
  $\dd J_{\nu,a} / \dd \alpha_{\nu,\rmn{3D}}$, where each point
  represents a radio relic in cluster g72a. Shown are {\em observable}
  relics (red crosses), {\em theoretical} relics (blue circles) and
  the large relic is emphasized by the bold cross and the filled
  circle. The dashed line is the theoretical Mach number computed from
  directly from the $\alpha_{\nu,\mathrm{3D}}$ (cf. Appendix
  Eqn. \ref{eq:mach}).  The theoretical Mach number over-predicts the
  median of the weighted Mach number, due to the weighting of
  $\alpha_{\nu,\mathrm{3D}}$ by the skewed distribution function of
  the radio luminosity.}
\label{fig:machalpha}
\end{figure}

In the following we want to address possible biases with our method
and show that the discrepancy between the predicted values and the
average radial value of the WHIM can be explained by differences
between the calculated Mach numbers and the median of the weighted
Mach numbers in the radio relics (cf. Fig. \ref{fig:machalpha}).  We
find that the weighted Mach numbers have systematically lower values
compared to the theoretical expectation due to the skewed distribution
of the emissivity weighted $\alpha_{\nu,\mathrm{3D}}$. According to
the Rankine-Hugoniot jump conditions, systematically higher values of
the shock strength should over-estimate the jumps and hence
under-predict all the pre-shock quantities, which appears to be the
case (Fig.~\ref{fig:WHIM}).

\subsection{Dependence on dynamical state and cluster mass}
\begin{figure*}
\begin{center}
 \begin{minipage}[t]{0.495\textwidth}
   \centering{\it \large Observable parameters}
 \end{minipage}
 \hfill
 \begin{minipage}[t]{0.495\textwidth}
   \centering{\it \large Theoretical parameters}
 \end{minipage}
\resizebox{0.5\hsize}{!}{\includegraphics{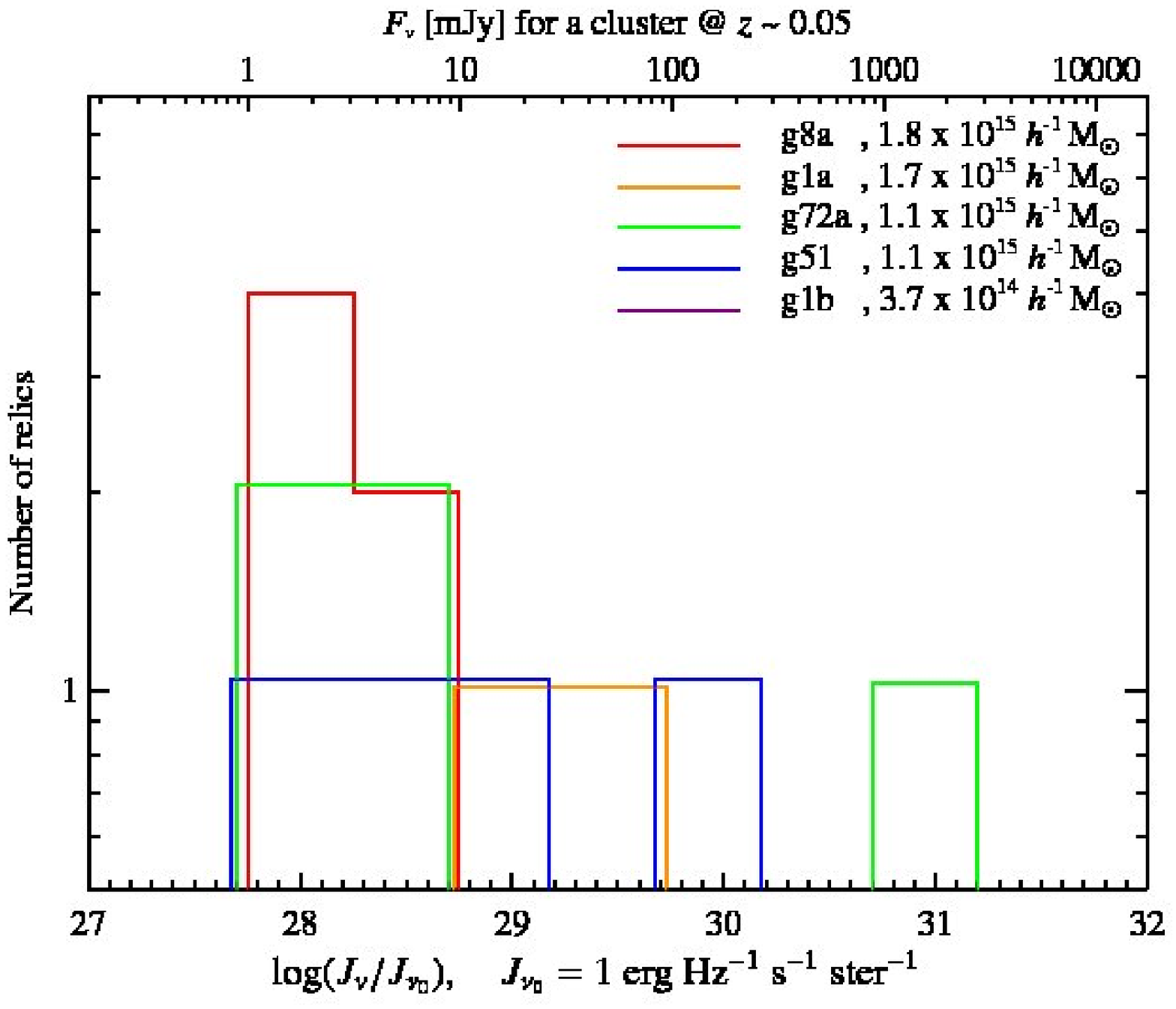}}%
\resizebox{0.5\hsize}{!}{\includegraphics{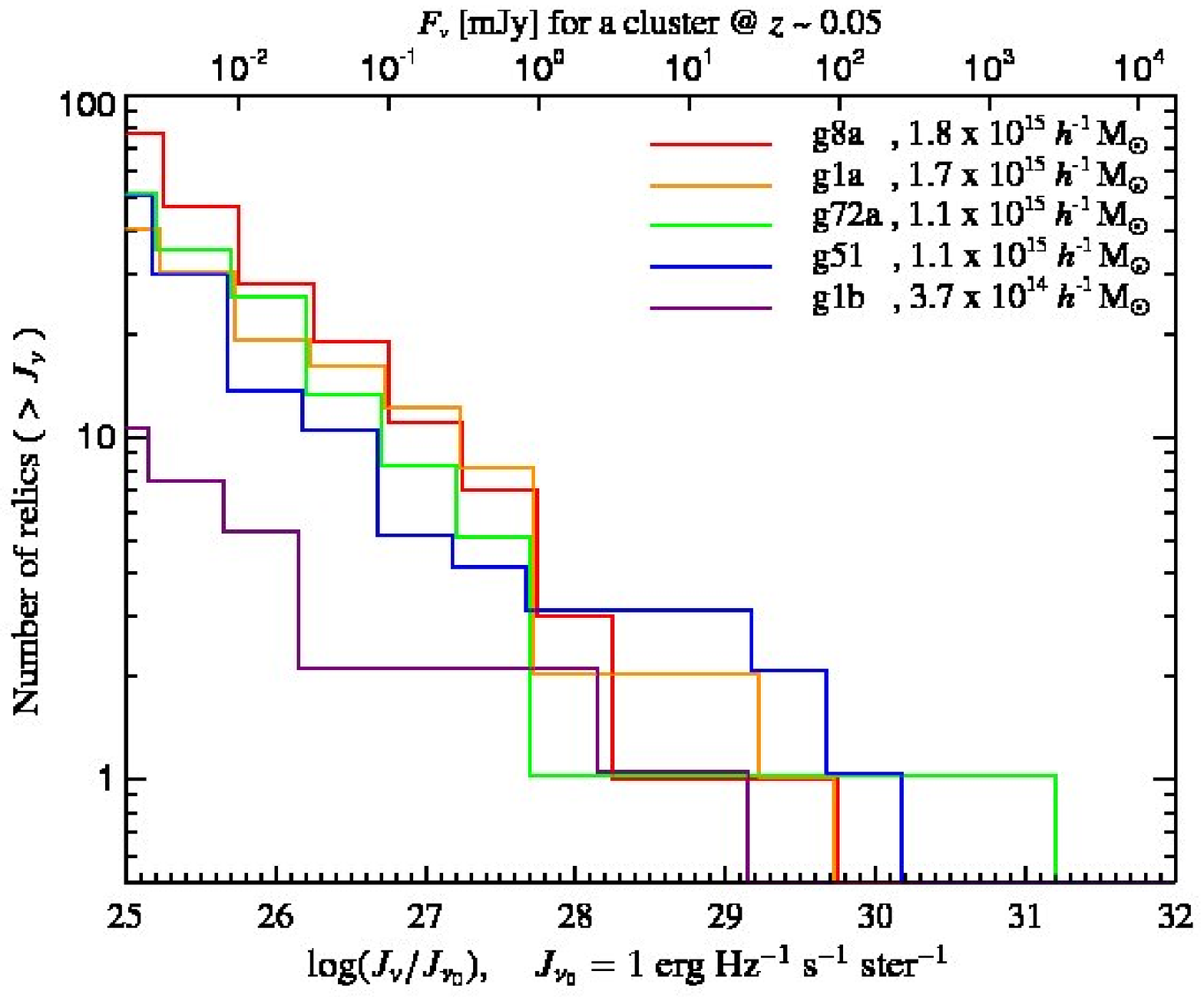}}\\
\resizebox{0.5\hsize}{!}{\includegraphics{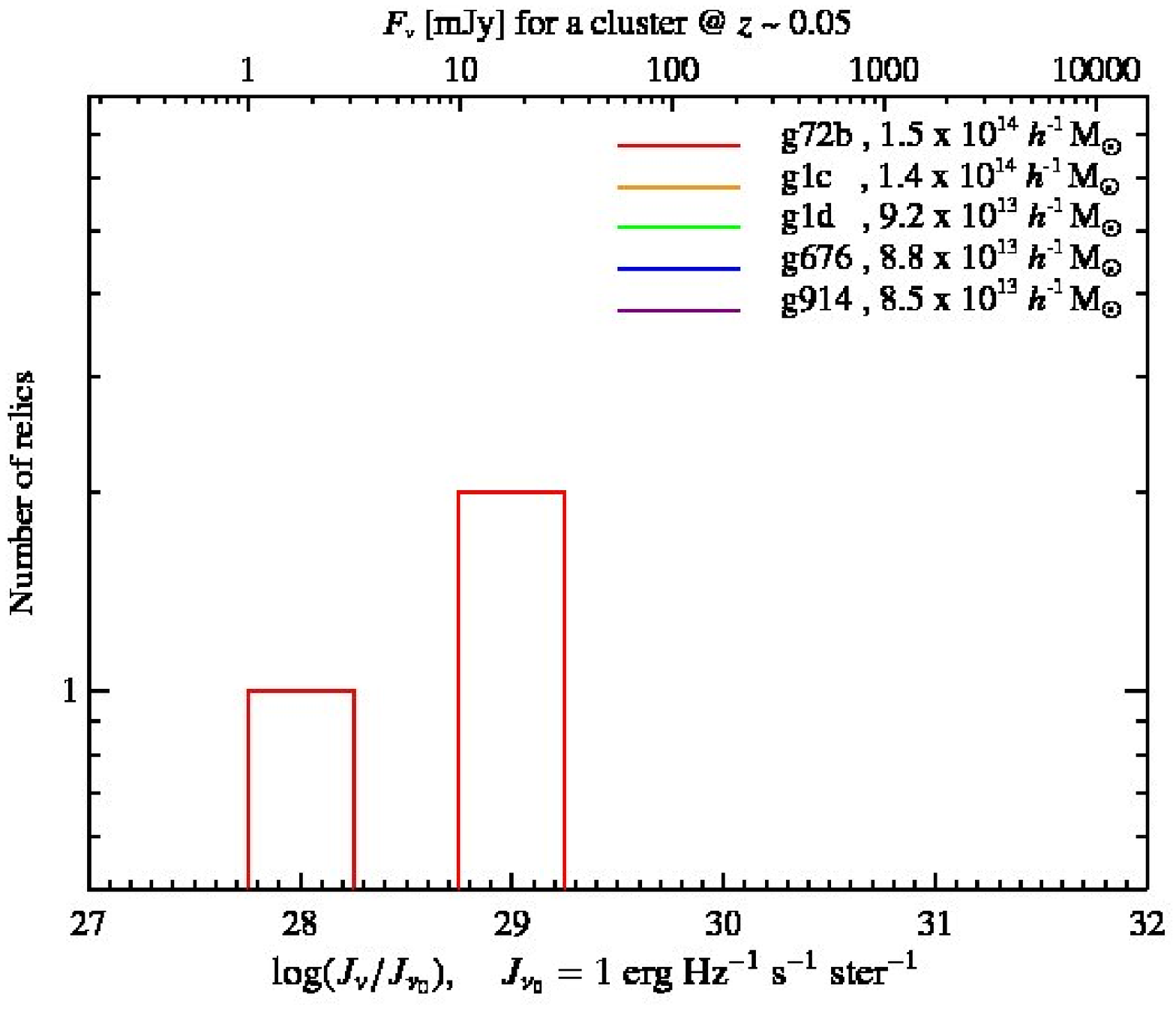}}%
\resizebox{0.5\hsize}{!}{\includegraphics{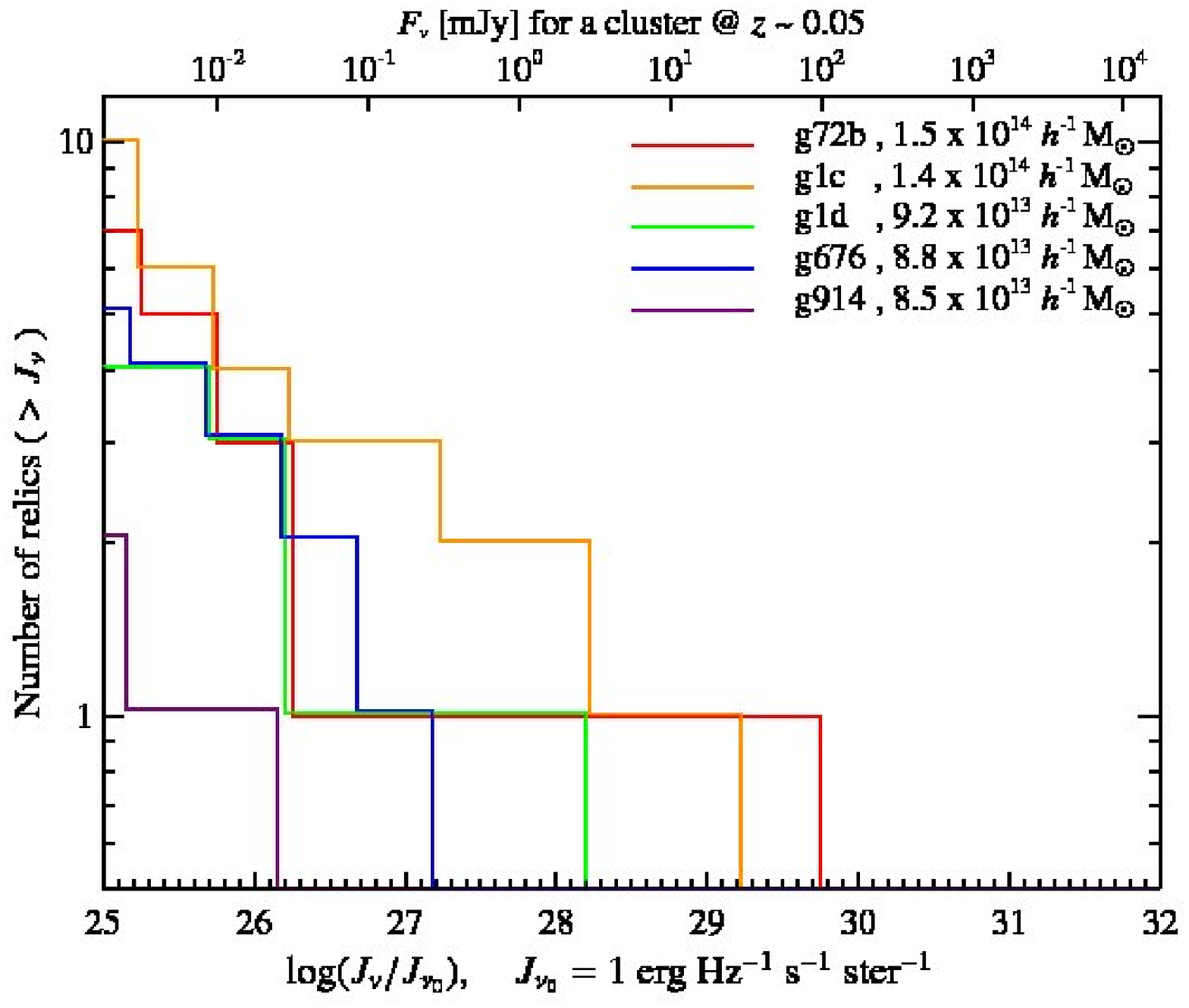}}\\
\end{center}
\caption{Luminosity functions for our sample of 10 clusters with the
  high mass clusters ($M_\rmn{vir} > 2 \times 10^{14}
  \,h^{-1}\rmn{M}_{\sun}$) in the upper panels and the low mass
  clusters in the lower panels for our standard magnetic model at
  150 MHz (see Table~\ref{tab:bpar}). The left panels contain relics
  found using the observable parameters and the right panels contain
  the theoretical parameters. The luminosity functions from the
  observable parameters show that more massive clusters have relics
  while low mass clusters have no relics (with g1b and g72b being the
  exceptions in both cases, respectively). The cumulative luminosity
  functions from the theoretical parameters show the trend for higher
  mass clusters to contain more relics and dynamical cluster stage
  modulates this effect notably, especially at low cluster masses. For
  instance compare the luminosity functions of the equal mass clusters
  g51 and g72a, the latter of which is a post-merging cluster.}
\label{fig:lfs_all}
\end{figure*}

\begin{figure}
  \resizebox{\hsize}{!}{\includegraphics{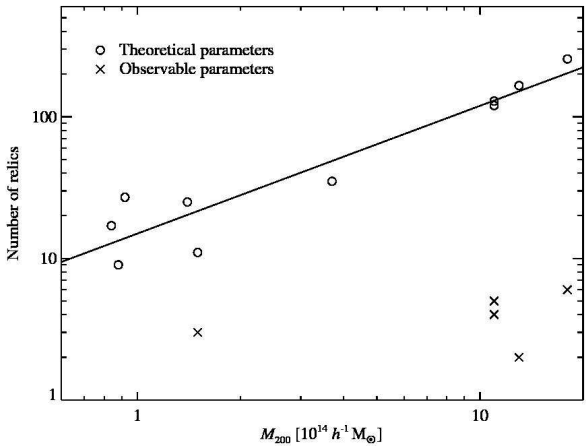}}
  \caption{Relation between the number of radio relics within a galaxy cluster
    and the cluster mass, where each point represents one of our simulated
    clusters and the line represent the best fit power-law that scales as
    $M_{200}^{0.9}$. There is a trend for higher mass clusters to contain more
    relics.}
\label{fig:mass_relic}
\end{figure}

We study the distribution of radio relics for the entire galaxy
cluster sample, which shows a variety of both dynamical states
(ranging from merging to cool core clusters) and masses (a range of
almost two orders of magnitude).  In Fig.~\ref{fig:lfs_all}, we
investigate how cluster mass and dynamical state depend on the relic
luminosity function. In the case of our theoretical parameter space,
more massive galaxy clusters clearly have more radio relics than the
lower mass clusters with a power-law scaling of $M_{200}^{0.9}$
(Fig. \ref{fig:mass_relic}). For the current observational
capabilities, we predict that only the most massive clusters should
have a significant sample of radio relics.

Ideally, one would like to directly compare clusters with the purpose of using
relic number statistics as a mass proxy. However, the luminosity functions have
another trend, which causes the scatter in the relationship between clusters
mass and total number of relics per cluster. This trend relates the cluster's
dynamical state to the luminosity of their brighter relics. The clusters g72a
and g51 have the same virial mass $1.1\times 10^{15}$ M$_{\sun}$, but g51 is a
relaxed cool core clusters in contrast to the active state of g72a. One can see
the two most luminous radio relics of g72a are an order of magnitude brighter
than any of g51's relics. Furthermore, the total amount of relics of g72a is
greater than that of g51. Merging clusters inherently have more shocks which
yields to more high-energy CRes and magnetic field amplification resulting in
more radio relics. This trend is even more severe for galaxy clusters of smaller
mass (Fig \ref{fig:lfs_all}). Our results show a larger probability of observing
a relic in a more massive cluster that is dynamically active. This dependence on
mass and dynamical state offers a possible explanation for why all current
observed radio relics are in massive merging clusters. They are expected to be
the brightest of a dimmer population of radio relics.

\section{Discussion}
\label{sec:discussion}

\subsection{Comparison with previous theoretical work}

Previously, there has been analytical work \citep{2004ApJ...617..281K} and
pioneering cosmological simulations \citep{2001ApJ...562..233M,
  2000ApJ...542..608M} on studying cluster synchrotron emission from
shock-accelerated electrons. The latter authors simulated the non-thermal
cluster emission by numerically modelling discretised cosmic ray (CR) energy
spectra on top of Eulerian grid-based cosmological simulations. Their approach
neglected the hydrodynamic pressure of the CR proton component, was quite
limited in its adaptive resolution capability, and neglected dissipative gas
physics including radiative cooling, star formation, and supernova feedback.  To
allow studies of the dynamical effects of CR protons in radiatively cooling
galactic and cluster environments, a CR proton formalism was developed that is
based on smoothed particle hydrodynamical representation of the equations of
motion \citep{2006MNRAS.367..113P, 2007A&A...473...41E,
  2008A&A...481...33J}. The emphasis is given to the dynamical impact of CR
protons on hydrodynamics, while simultaneously allowing for the important CR
proton injection and loss processes in a cosmological
setting. \citet{2008MNRAS.tmp..277P} modelled the CR electron components due to
shock acceleration as well as those being produced in hadronic CR proton
interaction with ambient gas protons.  Using this formalism and modelling,
\citet{2007MNRAS.378..385P, 2008MNRAS.tmp..277P} and \citet{2008MNRAS.tmp..278P}
coherently studied the non-thermal cluster emission, the cosmic ray pressure
component, and its implications for thermal cluster observables such as the
X-ray emission and the Sunyaev-Zel'dovich effect.  The focus of
\citet{2001ApJ...562..233M} was on primary CRe synchrotron emission from galaxy
clusters as a whole, whereas we focused in this work on the emission from
individual relics in detail to study the synchrotron observables and how they
are sensitive to the large scale magnetic properties. To summarize, we have
improved on past work in the observable predictions, simulations and
observational understanding of non-thermal emission from primary accelerated
CRe.

Additionally, we point out that IC and $\gamma$-ray emission are
alternate ways to study structure formation shocks. This was proposed
in analytical work by \citet{2000Natur.405..156L} and in simulations
\citep{2003MNRAS.342.1009M,2003ApJ...585..128K,2007ApJ...667L...1M,2008MNRAS.tmp..277P,2008MNRAS.tmp..278P}. We
are optimistic that high energy $\gamma$-ray experiments, such as the
Fermi $\gamma$-ray space telescope (formely GLAST) and future imaging
air \v{C}erenkov telescopes, will aid in further developing the
picture of non-thermal emission at structure formation shocks.

\subsection{Assumptions}

In our attempt to model diffuse radio relic emission from galaxy
clusters, we have made several simplifying assumptions. (1) We assume
the modified thermal leakage model fully describes the process of
diffusive shock acceleration and did not vary the parameters
associated with it. The described observations allow one to test the
self-consistency of this hypothesis, and might finally allow
improvement of our
knowledge about diffusive shock acceleration in high-$\beta$ plasmas.
(2) We neglect at this point the modifications of this simple model
due to non-linear shock acceleration, as well as cosmic ray modified
shocks, and postpone their study until future work.  (3) We also neglect
re-acceleration of mildly relativistic electrons that have been
injected in the past either by formation shocks or other sources such
as AGN.  (4) We use a simple parametrization for the magnetic
field. There are indications that the main characteristics of this
model are realized in clusters on average
\citep{2005A&A...434...67V,2004A&A...424..429M}. Future
work has to be dedicated to study the distribution of magnetic fields
that follow the magneto-hydrodynamics in radiative simulations. (5) We
are solving for a steady-state spectrum of the electron population and
are not sensitive to spectral aging processes across the relic as they
may have been found recently by \citet{2008A&A...486..347G}. (6) In
our model, we assume the thermal reservoir to be the source of
electrons.  (7) In our analysis, we only consider the rotation measure
signal from the line-of-sight integration of the density weighted
parallel magnetic field. We explicitly neglect possible contributions
from magnetic field amplifications due to post-shock turbulence local
to the shock wave \citep{2006ApJ...652.1246V,
2008Sci...320..909R}. These questions are beyond the scope of this
work and will be studied elsewhere.  (8) The mass contained within the
relics is not a physically relevant quantity and suffers from the
finite resolution of the SPH technique at the dilute shocks in the
virial regions of clusters and beyond. The radio luminosity, however,
is a robust prediction within a given acceleration model since it
reflects conserved quantities such as energy and mass across the
shock.

\section{Conclusions}
\label{sec:conclusions}

The intermittency and inhomogeneous nature of structure formation
shocks are characterized by a highly non-Gaussian distribution
function.  This requires numerical simulations to study the implied
non-universality of the induced radio relic (or gischt) emission. It
is hard to conceive of an adequate analytical approach to this
problem. Observing the polarized emission of a sample of relics at
different frequencies enables us to gain insight into the
non-equilibrium processes at work -- in particular into the interplay
of large scale magnetic fields and structure formation shocks.  The
relevant observables of the relics include morphology, spectral shape,
relic luminosity function and Faraday rotation measure. The
theoretical implications of radio relic observables are as follows:

We model the shock acceleration of electrons at formation shocks and find that
the {\em morphology} of radio relics unambiguously characterises the underlying
structure of dissipating shock waves (Fig.~\ref{fig:machnumbers}). The
resulting simulated relics are very similar to the observed relics and thus
support our hypothesis.  Their positions identify regions that are not in
equilibrium and where the electron and ion temperatures are expected to show
strong deviations due to the comparatively long Coulomb mean free path that
governs their equilibration process.

 The {\em relic luminosity function} is sensitive to the
combination of normalization and scaling properties of the magnetic
field with thermal energy density as well as the electron acceleration
efficiency, the mass and dynamical state of a cluster.  Thus, it can
provide hints about the processes that generate these large scale
fields and can help to disentangle the dominant transport processes
which include effects from magnetic flux freezing and growth by
turbulent dynamos.


The {\em rotation measure (RM) map} is sensitive to the line-of-sight
integrated magnetic field. From the mean and variance of RM maps,
we can infer the location of the formation shock with respect to
the cluster center as the variance increases as a function of
integration length (Eqn.~\ref{eq:RMsquared3}). This helps in
constraining the geometry of the merger. Deprojecting the 2D RM power
spectrum enables one to measure the 3D magnetic power spectrum, under
the assumption that the behaviour of the electron density along the
line of sight can be obtained from X-ray measurements. The peak of the
3D power spectrum yields the total magnetic energy and the magnetic
coherence length $\lambda_B$. Performing this procedure for different
relics or for different regions of one large relic allows us to
estimate the variance of $\lambda_B$ across the cluster and might
possibly tell us about the nature of MHD turbulence.  We found that
the correlation between $n_\e$ and $B$ biases the {\em rms} magnetic
field strength derived from RM maps high if this is not taken into
account; we note that most works have done so. This correlation should
be a natural consequence of MHD effects such as flux freezing. If
systematic errors associated with RM studies are smaller than
statistical ones, we find that measurements of the small scale slope
of the RM power spectrum are accurate enough to differentiate between
Kolmogorov ($k^{-5/3}$) and Burgers ($k^{-2}$) turbulence spectra. The
interpretation of these slopes is however not straight forward and
needs to account for an additional flattening due to small scale
fluctuations in $n_{\mathrm{e}}$.

The {\em spectral index} of a power law spectrum of a radio relic is a
measure of the shock strength of that relic. The median spectral index
of a distribution of relics probes a distribution of virializing
shocks and can give an indication of CR proton injection. This is of
particular relevance for questions concerning the pressure
contribution of non-thermal components and enables comparison with
predictions of hydrodynamical simulations.  The shape of the spectrum
is sensitive to the acceleration mechanism of the relativistic
electrons and to their cooling processes. The variation of the
spectrum over the relic allows one to infer in situ magnetic field
strengths by comparing the synchrotron and IC cooling times to the
advection time downstream provided that the magnetic energy density is
not much smaller than the CMB energy density. This might constrain
models for the magnetic amplification at shocks in high beta
plasmas. A radio relic's luminosity is roughly correlated with the
shock strength. Thus, it is favourable to look for radio relics in
large, dynamically disturbed clusters, or use relic detections as a
proxy for dynamical activity of clusters \citep{2001A&A...378..408S}.

We demonstrated that the combination of {\em the relic spectral index
  with deprojected X-ray and SZ profiles} allows one to indirectly
  infer upper limits on the density and temperature of the warm
  hot intergalactic medium. Simulations show that the WHIM is not
  uniform, rather it is characterized by highly inhomogeneous
  structure that shows intermittent accretion events that are
  channeled mostly through filaments.

We predict that there will be a large sample of polarized radio
relics from a considerable number clusters in the near future. This
sample should allow one to constrain macroscopic model parameters,
which are expected to be higly non-Gaussian, using a joint analysis
method on the radio observables. For example, the combination of the
relic luminosity function, RM power spectra, X-ray and SZ measurements
should constrain the models of diffusive shock acceleration and large
scale magnetic fields. Future work will address the details of this
procedure.

\section*{Acknowledgments}
It is a pleasure to thank Volker Springel for providing us with
initial numerical algorithms on which we based parts of our
post-processing and for carefully reading the manuscript. We would
also like to thank the referee for an insightful and helpful report.

\bibliography{bibtex/chp}
\bibliographystyle{mn2e}

\appendix

\section{Interpolating and projecting SPH quantities}

In the course of this work we are required to interpolate our Lagrangian energy
density distribution as given by SPH on a 3D grid. We remind the reader that
the SPH smoothing kernel of an SPH particle $a$,
$W(|\vecbf{r}-\vecbf{r}_a|,h_a)$, is given by Eqn.~A.1 of
\citet{2001NewA....6...79S}. It is normalized in the continuum such that $\int
W(r,h) \dd^3 \vecbf{r} \equiv 1$.  A scalar field $x(\vecbf{r})$\footnote{We
  note that in general, $x$ has to be a thermodynamic extensive volume density
  such that the product $x\, M / \rho$ is extensive.} is interpolated onto a 3D
grid cell at $\vecbf{r}_{ijk}$ by the product of itself with the specific
volume $M_{a} / \rho_{a}$ of the gas particles over a comoving cube,
\begin{equation}
x(\vecbf{r}_{ijk}) = \frac{1}{L^{3}_{\mathrm{pix}}} 
\sum_a x_{a}\frac{M_{a}}{\rho_{a}} \W_{a,ijk}(|\vecbf{r}_{ijk} - \vecbf{r}_{a}|, h_a),
\label{eq:3dproj}
\end{equation}
where $L^{3}_{\mathrm{pix}}$ is the comoving volume of the grid cell and we define
the normalized 3D smoothing kernel of SPH particle $a$ at the grid position
$\vecbf{r}_{ijk}$ by
\begin{equation}
  \label{eq:normalized_kernel}
  \W(|\vecbf{r}_{ijk}-\vecbf{r}_a|,h_a) =
  \frac{W(|\vecbf{r}_{ijk}-\vecbf{r}_a|,h_a)}{\sum_{ijk} W(|\vecbf{r}_{ijk}-\vecbf{r}_a|,h_a)}.
\end{equation}
We note that the normalized interpolation conserves the interpolated quantity
strictly without any further requirement on the grid size.

Similarly, we employ the method of normalized projection of a three dimensional
SPH scalar fields $x(\vecbf{r})$ to perform projection integrals yielding the
quantity $X(\vecbf{r}_{\perp})$. In analogy to Eqn.~\ref{eq:3dproj} we obtain
\begin{equation}
X(\vecbf{r}_{\perp,ij}) = \frac{1}{L^{2}_{\mathrm{pix}}} 
\sum_a x_{a}\frac{M_{a}}{\rho_{a}} 
\mathcal{Y}_{a,ij}(|\vecbf{r}_{\perp,ij} - \vecbf{r}_{a}|,h_a),
\label{eq:2dproj}
\end{equation}
where $L^{2}_{\mathrm{pix}}$ is the comoving area of the pixel and the
normalized 2D projected smoothing kernel of SPH particle $a$ at the grid
position $\vecbf{r}_{\perp,ij}$ derives from the projected SPH kernel
$Y(|\vecbf{r}_{\perp,ij}|,h_a)$ and is given by
\begin{equation}
  \label{eq:normalized_kernel2}
  \mathcal{Y}(|\vecbf{r}_{\perp, ij}-\vecbf{r}_a|,h_a) =
  \frac{Y(|\vecbf{r}_{\perp,ij}-\vecbf{r}_a|,h_a)}{\sum_{ijk} Y(|\vecbf{r}_{\perp,ij}-\vecbf{r}_a|,h_a)}.
\end{equation}

\section{Theoretical emission threshold}
\label{sec:Jontest}

\begin{figure}
  \resizebox{\hsize}{!}{\includegraphics{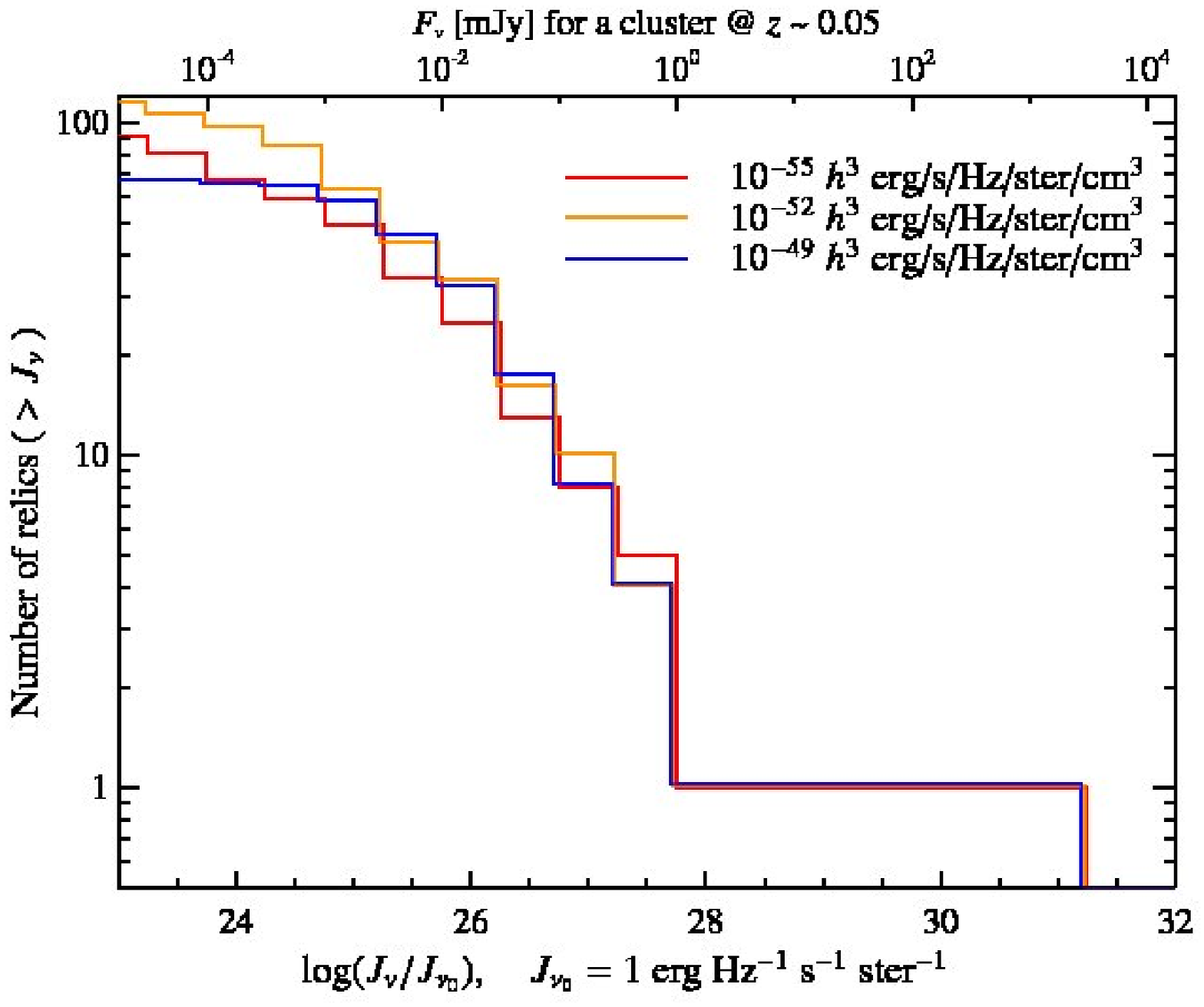}}
  \caption{Cumulative luminosity functions for different values of the emission
    cutoff. This shows the robustness of our predictions for future instrument
    capabilities.}
\label{fig:ET}
\end{figure}

The emission threshold for the {\em observable parameters} and {\em theoretical
  parameters} differ by 12 orders of magnitude. This dynamic range is beyond the
ability of any future telescope on the horizon. Varying the emission threshold of
our {\em theoretical parameters} by six orders of magnitudes only very
weakly affects our results. In particular, we show in Fig.~\ref{fig:ET} that such
a dramatic variation has only little influence on the high-end of the radio relic
luminosity function.

\section{Rankine-Hugoniot conditions}
\label{sect:RHjump}

The three dimensional spectral index can be transformed into a Mach number
($\mathcal{M}$) \citep{2007A&A...473...41E}, if one assumes an ideal fluid that
is characterized a a single adiabatic index $\gamma$,
\begin{equation}
\mathcal{M} = 
\sqrt{\frac{4(1 + \alpha_{\nu,\mathrm{3D}})}{1 + 4\alpha_{\nu,\mathrm{3D}} - 3\gamma}}.
\label{eq:mach}
\end{equation}
Under these conditions, the well-known Rankine-Hugoniot jump conditions allow to
relate the hydrodynamic post-shock quantities (denoted with a subscript 2) to
the pre-shock quantities (denoted with a subscript 1),
\begin{eqnarray}
\frac{\rho_1}{\rho_2} &=&
  \frac{(\gamma - 1)\mathcal{M}^2 + 2}{(\gamma + 1)\mathcal{M}^2},
\label{eq:densjump}\\
\frac{T_1}{T_2} &=&
  \frac{(\gamma + 1)^2\mathcal{M}^2 + 2}
   {\left[ 2\gamma\mathcal{M}^2 -(\gamma - 1) \right]
     \left[(\gamma - 1)\mathcal{M}^2 + 2\right]}, 
\label{eq:tempjump}\\
\frac{P_1}{P_2}&=&
 \frac{\gamma + 1}{ 2\gamma\mathcal{M}^2 -(\gamma - 1)}. 
\label{eq:pthjump}
\end{eqnarray}
Phenomenologically, we show in Fig.~\ref{fig:machalpha} that Eqn.~\ref{eq:mach}
under-predicts the average $\bra\mathcal{M}\ket$ if one were to infer
$\mathcal{M}$ from spectral index maps.  This translates into an upper limit
for the predicted pre-shock density.  In the case of temperature and pressure,
the under-prediction of the average $\bra\mathcal{M}\ket$ leads to an
over-estimation of the pre-shock values which translates into lower limits for
temperature and pressure (Eqns.~\ref{eq:tempjump} and \ref{eq:pthjump}).

\bsp

\label{lastpage}

\end{document}